\documentclass[11pt,a4paper]{article}

\usepackage[utf8]{inputenc}
\usepackage[T1]{fontenc}
\usepackage{charter}

\usepackage[margin=2.5cm]{geometry}
\usepackage{amsmath,amssymb,amsthm,mathtools}
\usepackage{bm}
\usepackage{booktabs}
\usepackage{enumitem}
\usepackage{microtype}
\usepackage{xcolor}
\usepackage{graphicx}
\usepackage[colorlinks=true,linkcolor=blue!50!black,citecolor=blue!50!black,urlcolor=blue!50!black]{hyperref}
\hypersetup{pdftitle={Energy-preserving Bayesian model reduction: closed-form structured posteriors and probabilistic stability certificates for quadratic operator inference},
  pdfauthor={Joao Victor Galvao da Mata, Mengwu Guo}}
\usepackage{tikz-cd}
\usetikzlibrary{arrows.meta,positioning,fit}
\usepackage{cleveref}
\usepackage{algorithm}
\usepackage{algpseudocode}
\usepackage{authblk}
\numberwithin{equation}{section}
\theoremstyle{plain}
\newtheorem{theorem}{Theorem}[section]
\newtheorem{proposition}[theorem]{Proposition}
\newtheorem{lemma}[theorem]{Lemma}
\newtheorem{corollary}[theorem]{Corollary}
\theoremstyle{definition}
\newtheorem{definition}[theorem]{Definition}

\theoremstyle{remark}
\newtheorem{remark}[theorem]{Remark}

\crefname{assumption}{Assumption}{Assumptions}
\algrenewcommand\algorithmicrequire{\textbf{Input:}}
\algrenewcommand\algorithmicensure{\textbf{Output:}}

\newcommand{\R}{\mathbb{R}}
\newcommand{\Prob}{\mathbb{P}}
\newcommand{\E}{\mathbb{E}}
\newcommand{\Nor}{\mathcal{N}}
\newcommand{\GP}{\mathcal{GP}}
\newcommand{\T}{^{\mathsf{T}}}
\newcommand{\q}{\hat{\mathbf{q}}}
\newcommand{\qq}{\mathbf{q}}
\newcommand{\bu}{\mathbf{u}}
\newcommand{\m}{\mathbf{m}}
\newcommand{\z}{\mathbf{z}}
\newcommand{\s}{\mathbf{s}}
\newcommand{\y}{\mathbf{y}}
\newcommand{\Ohat}{\widehat{\mathbf{O}}}
\newcommand{\ohat}{\hat{\mathbf{o}}}
\newcommand{\Ahat}{\widehat{\mathbf{A}}}
\newcommand{\Hhat}{\widehat{\mathbf{H}}}
\newcommand{\Bhat}{\widehat{\mathbf{B}}}
\newcommand{\chat}{\hat{\mathbf{c}}}
\newcommand{\Dt}{\widetilde{\mathbf{D}}}
\newcommand{\zt}{\tilde{\mathbf{z}}}
\newcommand{\W}{\mathbf{W}}
\newcommand{\C}{\mathbf{C}}

\newcommand{\Tm}{\mathbf{T}}
\newcommand{\Pm}{\mathbf{P}}
\newcommand{\K}{\mathbf{K}}
\newcommand{\Lm}{\mathbf{L}}
\newcommand{\Id}{\mathbf{I}}
\newcommand{\V}{\mathbf{V}}

\newcommand{\Hm}{\mathbf{H}}
\newcommand{\diag}{\operatorname{diag}}
\newcommand{\bd}{\mathbf{d}}
\newcommand{\bth}{\bm\theta}
\newcommand{\bvp}{\bm\varphi}
\newcommand{\bmu}{\bm\mu}
\newcommand{\bSig}{\bm\Sigma}
\newcommand{\bGam}{\bm\Gamma}
\newcommand{\Phim}{\bm\Phi}
\newcommand{\bet}{\bm\eta}
\newcommand{\sym}{\operatorname{sym}}
\newcommand{\vecop}{\operatorname{vec}}
\newcommand{\lmax}{\lambda_{\max}}

\newcommand{\blkdiag}{\operatorname{blkdiag}}
\newcommand{\range}{\operatorname{range}}
\newcommand{\SEP}{\mathcal{H}_{\mathrm{EP}}}
\newcommand{\SSK}{\mathcal{A}_{\mathrm{sk}}}

\title{\bfseries Energy-preserving Bayesian model reduction: closed-form structured posteriors and probabilistic stability certificates for quadratic operator inference}
\author[1]{Jo\~ao Victor Galv\~ao da Mata\thanks{Corresponding author. E-mail: \texttt{joao\_victor.galvao\_da\_mata@math.lu.se}.}}
\author[1]{Mengwu Guo\thanks{E-mail: \texttt{mengwu.guo@math.lu.se}.}}
\affil[1]{Centre for Mathematical Sciences, Lund University, Box 118, SE-221 00 Lund, Sweden}
\date{}

\makeatletter
\let\@oldappendix\appendix
\renewcommand\appendix{%
  \@oldappendix
  \renewcommand\@seccntformat[1]{\csname ##1@cntformat\endcsname}%
  \@namedef{section@cntformat}{Appendix~\thesection:\quad}%
  \@namedef{subsection@cntformat}{\thesubsection\quad}%
  \@namedef{subsubsection@cntformat}{\thesubsubsection\quad}%
}
\makeatother

\begin{document}
\sloppy
\maketitle

\begin{abstract}
\noindent
Bayesian operator inference learns reduced-order models (ROMs) with polynomial
structure from state data and returns a closed-form Gaussian posterior over the
reduced operators. That posterior is unconstrained, so it gives positive
probability to operators with unbounded dynamics, and current practice controls
this only through finite-horizon simulation tests on a few samples. We show that
energy preservation of the quadratic operator is a set of linear equality
constraints, which can be imposed exactly inside the Bayesian formulation
without losing the closed form. In particular, the constrained posterior follows
from one Kalman-type update of the unconstrained one, and its samples are
obtained exactly, by the same correction applied to unconstrained draws. Under
this structure, global boundedness for all initial conditions and bounded inputs
reduces to a linear matrix condition on the symmetric part of the linear
operator. The simulation test is then replaced by a probabilistic stability
certificate: the posterior probability that the ROM is globally bounded,
estimated with exact confidence bounds; the posterior restricted to certified
models is then sampled exactly by rejection. For energy-conserving systems, the
linear part can be constrained to be skew-symmetric as well: every draw then grows at most linearly in time, and conserves energy exactly when unforced.

On a forced viscous Burgers’ equation, the Korteweg–de Vries equation and a
two-dimensional viscous Burgers’ equation, every sample of the constrained
posterior integrates to a bounded trajectory, where up to all unconstrained
draws diverge, and beyond the observation window the former is also the more accurate of the two.

\end{abstract}

\section{Introduction}
\label{sec:intro}

Many tasks in computational science require repeated simulations of the same
underlying system. Design optimization, parameter estimation, uncertainty
quantification, and real-time prediction may each demand thousands of
evaluations across different characterizing parameters, configurations, external inputs, initial or boundary conditions. When each evaluation
requires solving a large-scale model, these many-query tasks can be prohibitively
expensive. Reduced-order models (ROMs) address this challenge by approximating the
dynamics in a low-dimensional latent space, so that repeated predictions can be made
at a small fraction of the cost of the full-order model (FOM).

ROMs can be constructed intrusively or non-intrusively. Intrusive methods typically use
the governing equations or simulation code to project the FOM dynamics onto a
low-dimensional basis. Projection-based methods can often preserve mathematical and
physical properties of the original system in the reduced equations. Non-intrusive methods infer the
reduced dynamics from data, and therefore remain applicable when the
equations or solvers are unavailable, as is common with experimental data or
legacy software. This flexibility comes with a challenge: a fit to observed
trajectories does not, by itself, ensure that the inferred ROM retains
the mathematical and physical properties of the original system, such as
conservation laws, energy balance, or stability. These properties can be
especially important when a ROM is used beyond the conditions represented
in its training data.

Operator inference (OpInf)~\cite{PeherstorferWillcox2016,
KramerPeherstorferWillcox2024} is a non-intrusive approach that posits a
low-dimensional model form and estimates its operators by regression on
reduced-state snapshots and their time derivatives. In many applications,
however, snapshots are sparse in time, noisy, or both. Estimating derivatives
directly from such data can amplify measurement noise, while sparse temporal
sampling makes the state evolution harder to resolve. Bayesian OpInf
\cite{GuoMcQuarrieWillcox2022} places a probabilistic prior on the operators,
relating regularized least squares to a Gaussian posterior and providing a
natural representation of operator uncertainty. GP-BayesOpInf
\cite{McQuarrie2025} extends this idea by fitting Gaussian processes (GPs) to
the reduced coordinates. The GP regression smooths noisy observations and
provides analytic derivative estimates together with their uncertainty; these
quantities lead to a generalized least-squares fit and a closed-form posterior
over the reduced operators. The resulting uncertainty can then be propagated
to the ROM predictions, an essential part of assessing reliability when data
are limited and predictions extend beyond the observed trajectories.

Stability is a further requirement for a useful ROM: a surrogate whose
trajectories grow without bound may not provide reliable long-time or
out-of-training predictions. This issue is acute for quadratic models, whose
nonlinear terms can cause finite-time blow-up, and for projection-based models,
which can lose stability when dissipative modes are truncated
\cite{SchlegelNoack2015}. It is also structural in a Bayesian formulation. A
non-degenerate Gaussian posterior has full support over the operator
coefficients, so choosing its hyperparameters alone cannot rule out unstable
operators. In current practice, prior hyperparameters may be screened by
simulating a small number of posterior draws and rejecting values for which a
trajectory exceeds a chosen amplitude over a finite time horizon
\cite{McQuarrie2025,McQuarrie2021}. Such screening can be practically useful,
but a finite set of trajectories on a finite horizon does not certify global
boundedness, and stronger prior regularization also reduces posterior
uncertainty. Nor does this procedure carry a stability guarantee to new inputs,
initial conditions, or subsequent data updates.

This paper studies Bayesian learning of quadratic ROMs with a focus on
stability. A key structure is energy preservation: the quadratic part
redistributes energy among the reduced coordinates without creating or
dissipating net energy. For many convection-type systems, this property
follows from the energy balance of the full-order equations and is inherited
under Galerkin projection when the discretization and projection are
compatible. For data-driven models, energy preservation can be imposed through
linear equality constraints on the quadratic-operator coefficients.
Deterministic OpInf methods have already exploited this structure. Koike and
Qian~\cite{KoikeQian2024} formulate the constrained least-squares problem, give
the independent constraint count, and discuss inheritance under projection;
Gkimisis et al.~\cite{EPrepr2025} prove that every energy-preserving quadratic
operator has an equivalent representation with skew-symmetric blocks and use
it in a sequential least-squares method; Mou~\cite{Mou2026} imposes the same
constraints inside a maximum-likelihood fit of a stochastic model, estimating a
diffusion amplitude jointly with the quadratic-linear drift and using an
upper-triangular parametrization to remove the non-identifiability of the full
quadratic tensor.

Energy preservation is not the only structure that can be built into operator
inference. When the full-order system is canonically
Hamiltonian, Hamiltonian OpInf learns a reduced Hamiltonian in symplectic
coordinates, so that the inferred model is Hamiltonian by construction and
stays accurate well beyond the training interval~\cite{SharmaWangKramer2022};
related constructions preserve Lagrangian and multi-symplectic
structure~\cite{Yildiz2025}. Where no such geometric structure is available,
stability can be promoted by physics-informed regularization~\cite{Sawant2023}
or enforced exactly by parametrizing the operators so that only stable models
can be represented~\cite{Goyal2025}. These approaches share our premise, that
structure known from the physics should be imposed on the learned operators
rather than hoped for, and differ in which structure is imposed and at what
price: a nonlinear parametrization, for instance, secures stability but
forfeits the linear-in-the-coefficients form on which a closed-form Gaussian
posterior depends.

Stabilization of data-driven ROMs is not specific to operator inference: the
same energy-preserving structure is promoted through a trapping-region
formulation in sparse identification~\cite{Kaptanoglu2021}, and stability is
imposed as a constraint on the learned operator in system
identification~\cite{LacyBernstein2003,Umenberger2018}. Stability can also be
improved after the model is learned, by filtering the reduced solution as it is
computed~\cite{Moore2026}. The constraint used here acts earlier: the model
class is restricted before any data are seen, so the guarantee is a property of
every operator in the class rather than of a particular trajectory.

Building on these developments, we
incorporate energy preservation into a Bayesian model and study the resulting
stability guarantees and uncertainty quantification.

The contributions of this work are as follows:
\begin{enumerate}[leftmargin=*]
  \item We collect and extend the stability theory needed for
  energy-preserving quadratic ROMs (\cref{sec:ep}). In particular, we show
  that energy preservation is necessary for the Euclidean energy to be a
  global Lyapunov function, and that it gives global exponential stability
  when the constant term and input vanish, as well as an explicit
  input-to-state estimate for bounded inputs, under a single dissipativity
  condition on the linear operator (\cref{prop:odd,thm:GES,thm:ISS}). A
  shifted energy extends the analysis to non-dissipative linear parts
  (\cref{thm:trap}), while weak energy preservation yields explicit local
  stability radii (\cref{thm:weak}). Previously known results are attributed
  in \cref{rem:attr-index,rem:attr-count} and before \cref{thm:GES}.
  \item We develop energy-preserving Bayesian operator inference
  (\cref{sec:bayes}). The constrained posterior has equivalent
  interpretations from three perspectives: a degenerate prior on the energy-preserving subspace, a
  noise-free pseudo-observation, and a constrained MAP estimate with its
  reduced Hessian (\cref{thm:equiv}). Computationally, it follows from a single Kalman-type
  update of the GP-BayesOpInf posterior and admits exact samples by a linear
  correction (\cref{prop:matheron}). We also give a soft constraint equivalent
  to an anisotropic prior with Tikhonov regularization, with hyperparameters selected
  by closed-form marginal likelihood (\cref{prop:soft,prop:evidence}); the
  proposed method is recovered as a limiting case.
  \item We define posterior probabilities of certifiable stability and
  provide exact confidence intervals based on inexpensive eigenvalue
  computations, together with exact sampling from the posterior restricted
  to certified models (\cref{sec:cert,prop:truncated}).
  \item For conservative systems, which neither create nor consume energy and
  for which no decay rate exists to certify, we constrain the linear part as
  well, asking that $\sym\Ahat=\mathbf{0}$ (\cref{sec:conservative}). This
  amounts to adding rows to the constraint matrix, while the closed-form posterior,
  the Kalman update and the exact sampling remain unchanged. In this case, every draw obeys a deterministic bound: growth at
  most linear in time, and exact conservation of the reduced energy when the
  constant term vanishes (\cref{thm:skew}).
  \item We give algorithms that incorporate these steps into GP-BayesOpInf
  with negligible additional cost, and analyze their complexity
  (\cref{sec:alg}).
\end{enumerate}

The stability analysis builds on the long-term boundedness and shifted trapping
regions of Schlegel and Noack~\cite{SchlegelNoack2015}, later used in the
analysis of data-driven quadratic models by Kaptanoglu et
al.~\cite{Kaptanoglu2021}; stability domains for quadratic-bilinear ROMs are
studied by Kramer~\cite{Kramer2021}. We do not claim the energy-preservation
index conditions, their count, or their inheritance under Galerkin projection
as new results; we state them with attribution because the Bayesian
construction relies on their precise form. Our contribution is a posterior
supported on the energy-preserving subspace, with a closed-form covariance and
exact samples that remain compatible with sequential updating, as well as
explicit stability estimates for sampled operators and a posterior
probability of certified stability with exact confidence bounds. To our
knowledge, existing work does not provide this combination of structured
operator posteriors and probabilistic stability certification.

\paragraph{Outline.} \Cref{sec:background} reviews proper orthogonal decomposition (POD), OpInf, and
GP-BayesOpInf. \Cref{sec:ep} develops the stability theory.
\Cref{sec:bayes} builds energy preservation into the Bayesian model.
\Cref{sec:cert} introduces probabilistic certificates.
\Cref{sec:conservative} treats conservative systems, where no decay rate exists
and the linear part is constrained instead. \Cref{sec:alg}
collects the algorithms. \Cref{sec:numerics} is reserved for numerical
experiments, and \cref{sec:conclusion} concludes. Proofs of auxiliary results
are in Appendix~\ref{app:proofs}.

\paragraph{Notation.} $\|\cdot\|$ is the Euclidean norm, or the induced
spectral norm for matrices. $\|\cdot\|_F$ is the Frobenius norm.
$\sym(\mathbf{X})=\tfrac12(\mathbf{X}+\mathbf{X}\T)$. $\lmax$ is the
largest eigenvalue of a symmetric matrix. $\otimes$ is the Kronecker
product. $\Nor(\bmu,\bSig)$ denotes a (possibly degenerate) Gaussian distribution. For a positive semidefinite $\bSig$, $\bSig^{1/2}$ is its symmetric square
root.

\section{Background}
\label{sec:background}

\subsection{POD and quadratic reduced-order models}

Consider a full-order model (FOM)
\begin{equation}
  \dot{\qq}=\mathbf{f}(\qq,\bu),\qquad \qq(t)\in\R^N,
  \label{eq:fom}
\end{equation}
where $\qq(t)\in\R^N$ is the state, typically a field discretized on a fine
spatial grid, so that $N$ is large; $\bu(t)\in\R^p$ collects the inputs, such
as forcing terms, boundary data or control signals; and
$\mathbf{f}:\R^N\times\R^p\to\R^N$ is the vector field of the governing
equations, which may be available only as a simulation code, or not at all.
Snapshots of $\qq$ are collected at representative times and operating
conditions $\bu$. Given a reference state $\bar\qq$ (e.g. the average of snapshots), proper orthogonal decomposition
(POD) applies a singular value decomposition to the centered snapshots and
selects the leading left singular vectors. These vectors span a low-dimensional
subspace that captures the predominant variation in the data. Let
$\V\in\R^{N\times r}$ contain the resulting orthonormal basis, where the size of the reduced basis $r\ll N$
and $\V\T\V=\Id_r$. The reduced-order solution is given by an approximation in such a low-dimensional subspace:
\begin{equation}
  \qq\approx\bar\qq+\V\q,
  \label{eq:pod}
\end{equation}
so $\q\in\R^r$ contains the reduced coordinates. If the FOM is accessible,
substituting this approximation into the governing equations and projecting the residual
onto the reduced basis gives the intrusive reduced dynamics
$\dot\q=\V\T\mathbf{f}(\bar\qq+\V\q,\bu)$. This is the Galerkin-projection
ROM. When the vector field is
quadratic in the state, or can be expressed in quadratic form after a suitable
variable transformation~\cite{Qian2020}, the reduced dynamics take the form
\begin{equation}
  \dot\q=\chat+\Ahat\q+\Hhat(\q\otimes\q)+\Bhat\bu(t),\qquad \q(0)=\q_0,
  \label{eq:rom}
\end{equation}
where $\chat\in\R^r$, $\Ahat\in\R^{r\times r}$,
$\Hhat\in\R^{r\times r^2}$, and $\Bhat\in\R^{r\times p}$. The constant
and linear terms include the effects of expanding about the reference state.
The same model form \eqref{eq:rom} can also be learned from data, without
constructing its operators by projecting the FOM.

To describe the quadratic term, write
$H_{ijk}:=\Hhat_{i,(j-1)r+k}$, so that
$[\Hhat(\q\otimes\q)]_i=\sum_{j,k}H_{ijk}\hat q_j\hat q_k$. Because the
products commute, we may assume without loss of generality that
\begin{equation}
  H_{ijk}=H_{ikj}\qquad\forall\,i,j,k,
  \label{eq:symH}
\end{equation}
and define the associated symmetric bilinear map by
$\mathcal H(\y,\z):=\Hhat(\y\otimes\z)$. Only the
$r(r+1)/2$ products $\hat q_j\hat q_k$ with $j\le k$ are distinct. We
therefore represent the quadratic term using compact coefficients $h_{i,jk}$:
\begin{equation}
  \bigl[\Hhat(\q\otimes\q)\bigr]_i
  =\sum_{j\le k}h_{i,jk}\,\hat q_j\hat q_k,
  \qquad i=1,\dots,r,
  \label{eq:compactH}
\end{equation}
where $h_{i,jk}$ is the coefficient of $\hat q_j\hat q_k$ in equation $i$.
Under \eqref{eq:symH}, $h_{i,jj}=H_{ijj}$ and $h_{i,jk}=2H_{ijk}$ for
$j<k$, since the two symmetric tensor entries multiply the same product. In
the operator matrix below, the quadratic block stores these compact
coefficients and therefore has $r(r+1)/2$ columns.

\subsection{Operator inference}

When the FOM equations or solver are unavailable, OpInf
learns the reduced operators from reduced-state data and estimates of their
time derivatives. For \eqref{eq:rom}, collect the constant, linear, compact
quadratic, and input coefficients in
$\Ohat=[\chat\ \Ahat\ \Hhat\ \Bhat]\in\R^{r\times d}$, where the
quadratic block is understood in the compact form just defined. Then the model
can be written as a linear function of its unknown operators evaluated at a
nonlinear feature vector:
\begin{equation}
  \dot\q=\Ohat\,\bd(\q,\bu),\qquad
  \bd(\q,\bu)=[1;\ \q;\ \q\,\hat\otimes\,\q;\ \bu]\in\R^d,
  \qquad d=1+r+\frac{r(r+1)}{2}+p,
  \label{eq:features}
\end{equation}
where $\hat\otimes$ lists only the distinct quadratic products. Given reduced
states $\q$ and derivative estimates for $\q$ at $m'$ time instants, form the data matrix
$\mathbf{D}\in\R^{m'\times d}$ with the corresponding feature vectors as its
rows. For each reduced coordinate, let $\z_i$ collect its derivative
estimates and let $\ohat_i\T$ denote row $i$ of $\Ohat$. Regularized OpInf
then solves the independent least-squares problems for all reduced modes:
\begin{equation}
  \min_{\ohat_i\in\R^d}\quad
  \|\mathbf{D}\ohat_i-\z_i\|^2+\|\bGam_i\ohat_i\|^2,
  \qquad i=1,\dots,r,
  \label{eq:opinf}
\end{equation}
where $\bGam_i$ contains the regularization weights~\cite{PeherstorferWillcox2016,
KramerPeherstorferWillcox2024}. The regression
is linear in the operators, but its quality depends on the reduced states and
derivative estimates. This dependence is problematic when snapshots are sparse
or noisy. Finite-difference estimates amplify observation noise, and widely or
irregularly spaced snapshots leave the state evolution between them
unresolved, so that the estimate is inaccurate even in the absence of noise.

\subsection{Bayesian operator inference with Gaussian processes}

Bayesian OpInf~\cite{GuoMcQuarrieWillcox2022} gives the regularization a
probabilistic interpretation by placing a Gaussian prior on each operator row.
GP-BayesOpInf~\cite{McQuarrie2025} also models each reduced coordinate as an
independent Gaussian process, $\hat q_i\sim\GP(0,\kappa_i)$, conditioned on
noisy observations $\y_i\in\R^m$ at $m$ times. The GP posterior smooths these
observations and provides state estimates $\tilde{\y}_i$ and analytic
derivative estimates $\zt_i$ on an estimation grid. It also quantifies the
uncertainty in the derivative estimates. Let $\W_i\in\R^{m'\times m'}$ be
the inverse of the derivative estimates' covariance matrix, and form $\Dt$ from the GP state estimates.
With the prior $\ohat_i\sim\Nor(\mathbf{0},\bGam_i^{-2})$,
$\bGam_i=\diag(\bm\gamma_i)$, the resulting Gaussian posterior of the reduced operator $\ohat_i$ is
\begin{equation}
  \ohat_i\mid\zt_i,\Dt,\W_i\sim\Nor(\bmu_i,\bSig_i),\qquad
  \bSig_i=\bigl(\Dt\T\W_i\Dt+\bGam_i^2\bigr)^{-1},\qquad
  \bmu_i=\bSig_i\Dt\T\W_i\zt_i.
  \label{eq:rowpost}
\end{equation}
Equivalently, this posterior follows from the working linear-Gaussian model
\begin{equation}
  \zt_i=\Dt\ohat_i+\mathbf{e}_i,\qquad
  \mathbf{e}_i\sim\Nor(\mathbf{0},\W_i^{-1}),\qquad
  \ohat_i\sim\Nor(\mathbf{0},\bGam_i^{-2}).
  \label{eq:linmodel}
\end{equation}

For a joint representation, stack the operator rows as
$\bth=[\ohat_1;\dots;\ohat_r]\in\R^n$, where $n=rd$. Since the prior and
working likelihood factor over the output coordinates, the posterior is
\begin{equation}
  \bth\mid\zt,\Dt,\W\sim\Nor(\bmu,\bSig),\qquad
  \bmu=[\bmu_1;\dots;\bmu_r],\qquad
  \bSig=\blkdiag(\bSig_1,\dots,\bSig_r),
  \label{eq:jointpost}
\end{equation}
which is equivalently the posterior under
$\zt=\Phim\bth+\mathbf{e}$, with
$\zt=[\zt_1;\dots;\zt_r]$, $\Phim=\Id_r\otimes\Dt$,
$\mathbf{e}\sim\Nor(\mathbf{0},\W^{-1})$, and
$\W=\blkdiag(\W_i)$. The prior is
$\bth\sim\Nor(\mathbf{0},\bSig_0)$, where
$\bSig_0=\blkdiag(\bGam_i^{-2})$. Thus GP-BayesOpInf yields a Gaussian
posterior over the reduced operators, which represents operator uncertainty
and serves as the starting point for the structured posterior in
\cref{sec:bayes}.

\subsection{Simulation-based stability screening}
\label{sec:current}

The covariance matrix $\bSig$ in \eqref{eq:jointpost} is positive definite, so the posterior
has full support over the operator coefficients. In the GP-BayesOpInf method
of~\cite{McQuarrie2025}, extending~\cite{McQuarrie2021}, the prior parameters
$\bm\gamma=(\bm\gamma_1,\dots,\bm\gamma_r)$, which collect the prior scales of
all operator rows and so determine $\bSig_0$, are selected using
posterior-sampled ROM trajectories: candidate
values are compared by the fit between the GP state estimates and the mean of
an ensemble of trajectories, and rejected if any trajectory exceeds a chosen
amplitude over the training interval. The reported experiments use $20$
posterior samples and set this threshold to five times the largest absolute GP
state estimate. This simulation-based screening is a finite-sample diagnostic,
not a stability certificate. The present work retains the regression model
\eqref{eq:rowpost} while changing the posterior support and the way stability
is assessed and reported.

\section{Energy preservation and stability of quadratic ROMs}
\label{sec:ep}

Throughout this section the reduced operators in \eqref{eq:rom} are fixed. We use the
reduced energy
\begin{equation}
  V(\q)=\tfrac12\|\q\|^2 ,
  \label{eq:V}
\end{equation}
which, for an orthonormal basis $\V$, equals $\tfrac12\|\qq-\bar\qq\|^2$: the
energy of the fluctuation about the reference state. Differentiating $V(\q(t))$
in time and substituting the right-hand side of \eqref{eq:rom} for $\dot\q$
gives the energy balance
\begin{equation}
  \dot V=\q\T\chat+\q\T\sym(\Ahat)\,\q+p(\q)+\q\T\Bhat\bu,
  \qquad p(\q):=\q\T\Hhat(\q\otimes\q)=\sum_{i,j,k}H_{ijk}\hat q_i\hat q_j\hat q_k ,
  \label{eq:Vdot}
\end{equation}
where we used $\q\T\Ahat\q=\q\T\sym(\Ahat)\q$. The four terms scale as
$\|\q\|$, $\|\q\|^2$, $\|\q\|^3$ and $\|\q\|$, respectively. For large states
the cubic form $p$ dominates.

\subsection{Energy-preserving quadratic operators}

\begin{definition}
\label{def:ep}
The quadratic operator $\Hhat$ is \emph{energy-preserving} (EP) if
$p(\q)=\q\T\Hhat(\q\otimes\q)=0$ for all $\q\in\R^r$. We denote by $\SEP$ the
set of EP operators.
\end{definition}

\begin{proposition}[Index form]
\label{prop:index}
Under \eqref{eq:symH}, $\Hhat\in\SEP$ if and only if
\begin{equation}
  H_{ijk}+H_{jik}+H_{kij}=0\qquad\forall\,i,j,k\in\{1,\dots,r\}.
  \label{eq:EPindex}
\end{equation}
\end{proposition}
\begin{proof}
See Appendix~\ref{app:index}.
\end{proof}

\begin{remark}[Attribution]
\label{rem:attr-index}
Condition \eqref{eq:EPindex} is classical. It underlies the boundedness analysis
of Galerkin models in~\cite{SchlegelNoack2015} and the trapping-region approach
of~\cite{Kaptanoglu2021}, and it is stated in exactly this index form, for
operator inference, by Koike and Qian~\cite[eq.~(9)]{KoikeQian2024}. We restate
it here to fix notation and because the proof in Appendix~\ref{app:index} is what makes
the sparsity structure in \cref{cor:compact} explicit.
\end{remark}

\begin{corollary}[Compact form, rank and dimension]
\label{cor:compact}
In terms of the compact coefficients $h_{i,jk}$ of \eqref{eq:compactH},
$\Hhat\in\SEP$ if and only if, for all $a$, all $c>a$ and all $b \in \mathopen{]}a,c\mathclose{[}$
(indices in $\{1,\dots,r\}$),
\begin{equation}
  h_{a,aa}=0,\qquad h_{a,ac}+h_{c,aa}=0,\qquad h_{a,cc}+h_{c,ac}=0,\qquad
  h_{a,bc}+h_{b,ac}+h_{c,ab}=0 .
  \label{eq:compactC}
\end{equation}
There is one constraint per cubic monomial, $n_c=\binom{r+2}{3}$ constraints in
total. Each compact coefficient $h_{i,jk}$ appears in exactly one of them (the
one for the monomial $\hat q_i\hat q_j\hat q_k$), so the constraints are linearly
independent. Writing them as $\C_H\vecop(\Hhat)=\mathbf{0}$, $\SEP$ is a linear
subspace of dimension $r\binom{r+1}{2}-\binom{r+2}{3}=(r^3-r)/3$.
\end{corollary}
\begin{proof}
Multiplying \eqref{eq:compactH} by $\hat q_i$ and summing gives
$\q\T\Hhat(\q\otimes\q)=\sum_i\sum_{j\le k}h_{i,jk}\hat q_i\hat q_j\hat q_k$. Now
collect the coefficient of each monomial: $\hat q_a^3$, $\hat q_a^2\hat q_c$,
$\hat q_a\hat q_c^2$ and $\hat q_a\hat q_b\hat q_c$, respectively. Each coefficient $h_{i,jk}$
multiplies the monomial $\hat q_i\hat q_j\hat q_k$ and no other, so no two
constraints involve a common coefficient: every column of $\C_H$ has exactly one
nonzero entry. Hence, if a linear combination of the rows vanishes, then for each
row $\ell$ and any column $j$ in which row $\ell$ is nonzero, the $j$-th entry of
the combination reduces to the single term $c_\ell[\C_H]_{\ell j}$, giving
$c_\ell=0$; the rows are therefore linearly independent. The count is
$r+2\binom r2+\binom r3=\binom{r+2}3$.
\end{proof}

\begin{remark}[Attribution and consistency]
\label{rem:attr-count}
The count $n_c=r(r+1)(r+2)/6$ of independent constraints, and the reduction to
the half-vectorized (compact) parametrization of $\Hhat$, are due to Koike and
Qian~\cite{KoikeQian2024}; the same constraint set and count, written in the
compact (upper-triangular) coordinates of \eqref{eq:compactC}, are obtained
independently by Mou~\cite[Prop.~3.3]{Mou2026}. What we use in addition is the \emph{sparsity
pattern} established in the proof: since each compact coefficient enters exactly
one constraint, $\C_H$ has one nonzero per column, its rows are orthogonal, and
$\C_H\C_H\T=\diag(\nu_1,\dots,\nu_{n_c})$ with $\nu_\ell\in\{1,2,3\}$ the number of
terms in constraint $\ell$. Orthonormalizing the rows, which \cref{sec:soft}
requires, is therefore a rescaling by $\nu_\ell^{-1/2}$ rather than a
factorization. The dimension $(r^3-r)/3$ is consistent with the
skew-symmetric representation of Gkimisis et al.~\cite{EPrepr2025}: their
representation uses $r$ skew blocks, i.e.\ $r^2(r-1)/2$ parameters, of which
$r^3/6-r^2/2+r/3$ are redundant, and
$r^2(r-1)/2-(r^3/6-r^2/2+r/3)=(r^3-r)/3$.
\end{remark}

The next result explains why EP is natural for projection-based ROMs. It is
classical and is stated, in the operator-inference context, in~\cite{KoikeQian2024};
we include it because \cref{sec:bayes} uses the precise form of the mean-flow
term.

\begin{proposition}[Galerkin projection preserves EP]
\label{prop:galerkin}
Let the FOM be $\dot\qq=\mathbf{A}_F\qq+\Hm_F(\qq\otimes\qq)+\mathbf{B}_F\bu$ with
$\qq\T\Hm_F(\qq\otimes\qq)=0$ for all $\qq$, and let $\V$ have orthonormal
columns, with $\Hm_F$ symmetrized as in \eqref{eq:symH}. Then the Galerkin ROM
for $\qq=\bar\qq+\V\q$ has
\begin{equation}
  \Hhat=\V\T\Hm_F(\V\otimes\V)\in\SEP,\qquad
  \Ahat=\V\T\bigl(\mathbf{A}_F+2\Hm_F(\bar\qq\otimes\Id_N)\bigr)\V,
  \label{eq:galAH}
\end{equation}
\begin{equation}
  \chat=\V\T\bigl(\mathbf{A}_F\bar\qq+\Hm_F(\bar\qq\otimes\bar\qq)\bigr),\qquad
  \Bhat=\V\T\mathbf{B}_F,
  \label{eq:galcB}
\end{equation}
where $\Hm_F(\bar\qq\otimes\Id_N)\in\R^{N\times N}$ denotes the matrix of the linear
map $\mathbf{z}\mapsto\Hm_F(\bar\qq\otimes\mathbf{z})$.
\end{proposition}
\begin{proof}
Since $\dot\qq=\V\dot\q$ and $\V\T\V=\Id_r$, it holds that
$\dot\q=\V\T\mathbf{f}(\bar\qq+\V\q,\bu)$. Expanding the quadratic term by bilinearity and using the symmetry of
$\Hm_F$,
\[
  \Hm_F\bigl((\bar\qq+\V\q)\otimes(\bar\qq+\V\q)\bigr)
  =\Hm_F(\bar\qq\otimes\bar\qq)+2\Hm_F(\bar\qq\otimes\Id_N)\V\q
   +\Hm_F(\V\otimes\V)(\q\otimes\q),
\]
where the last term uses $(\V\otimes\V)(\q\otimes\q)=(\V\q)\otimes(\V\q)$.
Collecting powers of $\q$ gives \eqref{eq:galAH}--\eqref{eq:galcB}. Finally,
$\q\T\Hhat(\q\otimes\q)=(\V\q)\T\Hm_F((\V\q)\otimes(\V\q))=0$ for all $\q$ as $\V\q\in\R^N$, so
$\Hhat\in\SEP$.
\end{proof}

One consequence is used later: by \eqref{eq:galAH} the mean-flow term
$2\Hm_F(\bar\qq\otimes\Id_N)$ enters $\Ahat$, so $\sym(\Ahat)$ can be
indefinite even when the physics is dissipative. This is why \cref{sec:shift}
shifts the energy.

The following proposition makes explicit an observation that is implicit in the
boundedness analyses of~\cite{SchlegelNoack2015,Kaptanoglu2021}: EP is not merely
convenient, it is the only property of a quadratic term compatible with
\eqref{eq:V} being a global Lyapunov function.

\begin{proposition}[Necessity of EP]
\label{prop:odd}
The cubic form $p$ is odd, i.e. $p(-\q)=-p(\q)$ for every $\q\in\R^r$. Consequently
\begin{enumerate}[label=(\roman*),leftmargin=*,nosep]
  \item $p(\q)\le0$ for all $\q\in\R^r$ if and only if $p\equiv0$, that is, if
  and only if $\Hhat\in\SEP$;
  \item if $\Hhat\notin\SEP$, then whatever the values of $\chat$, $\Ahat$ and
  $\Bhat$, and for every bounded input $\bu$, there are states of arbitrarily
  large norm at which $\dot V>0$.
\end{enumerate}
In particular, unless $\Hhat\in\SEP$, the energy \eqref{eq:V} cannot certify
global boundedness, no matter how the remaining operators are chosen.
\end{proposition}
\begin{proof}
Oddness is immediate from homogeneity of degree three. For (i), if $p\le0$
everywhere then $p(\q)\le0$ and $-p(\q)=p(-\q)\le0$, so $p\equiv0$, the
reverse implication is immediate. For (ii), by (i) there is $\q_0$ with $p(\q_0)>0$. The
cubic term of \eqref{eq:Vdot} grows with the cube of $\|\q\|$, while the
remaining terms grow at most quadratically since $\bu$ is bounded, so
$\dot V>0$ at all sufficiently large multiples of $\q_0$.
\end{proof}

\begin{corollary}
\label{cor:as}
If $\bth\sim\Nor(\bmu,\bSig)$ with $\bSig\succ0$, then $\Hhat\notin\SEP$ almost
surely. Hence, almost surely, no posterior sample of \eqref{eq:jointpost} admits
\eqref{eq:V} as a global Lyapunov function.
\end{corollary}
\begin{proof}
$\SEP$ is a proper linear subspace, so it has Lebesgue measure zero, and a
non-degenerate Gaussian is absolutely continuous.
\end{proof}

\subsection{Global stability under energy preservation}

If $\Hhat\in\SEP$, the cubic term in \eqref{eq:Vdot} vanishes identically:
\begin{equation}
  \dot V=\q\T\chat+\q\T\sym(\Ahat)\q+\q\T\Bhat\bu .
  \label{eq:VdotEP}
\end{equation}
The energy balance of the nonlinear ROM is then that of a linear system, and
the estimates below specialize the boundedness analysis
of~\cite{SchlegelNoack2015}; we state them in the input-to-state form of
\cref{thm:ISS}, which is what \cref{sec:cert} certifies and which does not
appear in that reference.

Only two quantities enter: the rate at which the quadratic form in
\eqref{eq:VdotEP} removes energy, and the amount the remaining terms can
supply. Define
\begin{equation}
  \alpha:=-\lmax\bigl(\sym(\Ahat)\bigr),\qquad
  K(\bar u):=\|\chat\|+\|\Bhat\|\,\bar u,\qquad \bar u:=\|\bu\|_{L^\infty}.
  \label{eq:alphaK}
\end{equation}
If $\alpha>0$, then $\sym(\Ahat)\prec\mathbf{0}$ and
$\q\T\sym(\Ahat)\q\le-\alpha\|\q\|^2$, so the dissipation grows with
$\|\q\|^2$, whereas $\q\T\chat+\q\T\Bhat\bu\le K(\bar u)\|\q\|$ grows only with
$\|\q\|$. The two balance at $\|\q\|=K(\bar u)/\alpha$, which is the radius
that appears in the bounds below.

\begin{theorem}[Global exponential stability]
\label{thm:GES}
If $\Hhat\in\SEP$, $\chat=\mathbf{0}$, $\bu\equiv\mathbf{0}$ and $\alpha>0$,
then $\|\q(t)\|\le e^{-\alpha t}\|\q_0\|$ for all $\q_0\in\R^r$ and $t\ge0$.
\end{theorem}
\begin{proof}
By \eqref{eq:VdotEP}, $\dot V\le-\alpha\|\q\|^2=-2\alpha V$. Hence
$V(t)\le e^{-2\alpha t}V(0)$ by the comparison lemma~\cite[Lemma~3.4]{Khalil2002}.
\end{proof}

\begin{theorem}[Global ultimate bound and input-to-state stability]
\label{thm:ISS}
Let $\Hhat\in\SEP$, $\alpha>0$ and $\bar u<\infty$, and set $R:=K(\bar u)/\alpha$.
Then for all $\q_0$ and $t\ge0$,
\begin{equation}
  \|\q(t)\|^2\le e^{-\alpha t}\|\q_0\|^2+R^2,
  \qquad
  \|\q(t)\|\le e^{-\alpha t/2}\|\q_0\|+\frac{\|\chat\|}{\alpha}+\frac{\|\Bhat\|}{\alpha}\,\bar u .
  \label{eq:ISS}
\end{equation}
Every closed ball $\{\|\q\|\le\rho\}$ with $\rho\ge R$ is forward invariant, and
$\limsup_{t\to\infty}\|\q(t)\|\le R$. In particular, \eqref{eq:rom} is
input-to-state stable with respect to $\bu$, with linear gain $\|\Bhat\|/\alpha$
and offset $\|\chat\|/\alpha$.
\end{theorem}
\begin{proof}
From \eqref{eq:VdotEP}, $\dot V\le-\alpha\|\q\|^2+K\|\q\|$ with $K=K(\bar u)$.
Young's inequality $K\|\q\|\le\tfrac\alpha2\|\q\|^2+\tfrac{K^2}{2\alpha}$ gives
$\dot V\le-\alpha V+K^2/(2\alpha)$. The comparison lemma yields
$V(t)\le e^{-\alpha t}V(0)+K^2/(2\alpha^2)$, which is the first bound. The second
follows from  the inequality $\sqrt{a+b}\le\sqrt a+\sqrt b$ for non-negative $a$, $b$. For invariance, note that
$\dot V\le\|\q\|(K-\alpha\|\q\|)<0$ whenever $\|\q\|>R$. The limsup follows
from the first bound.
\end{proof}

\begin{remark}
\label{rem:iss}
The terms $\q\T\chat$ and $\q\T\Bhat\bu$ are linear in $\q$ and hence
sign-indefinite. They need not be dissipative: they grow like $\|\q\|$,
whereas the dissipation grows like $\|\q\|^2$. The conclusion is therefore
boundedness, not convergence to the origin, which in general is not an
equilibrium when $\chat\neq\mathbf{0}$. The condition $\alpha>0$ is sufficient
but not necessary. A Hurwitz but non-normal $\Ahat$ may have indefinite
$\sym(\Ahat)$.
\end{remark}

\subsection{Non-dissipative linear parts: the shifted energy}
\label{sec:shift}

The condition $\alpha>0$ asks the energy to decrease about the origin of the
reduced coordinates, that is, about the snapshot mean $\bar\qq$. This is more
than the physics requires. By \eqref{eq:galAH} the mean-flow term enters
$\Ahat$, and POD truncation discards modes that carry part of the dissipation,
so $\sym(\Ahat)$ is often indefinite even when the ROM itself is well behaved;
in \cref{sec:numerics} this is the case for nearly every posterior draw. The
remedy is to measure the energy about another center.

Following~\cite{SchlegelNoack2015}, let $\m\in\R^r$ and write
$V_\m(\q)=\tfrac12\|\q-\m\|^2$. For $\s=\q-\m$, symmetry of
$\mathcal H$ gives
\begin{equation}
  \dot\s=\mathbf{g}(\m)+(\Ahat+\Lm_\m)\s+\Hhat(\s\otimes\s)+\Bhat\bu,
  \qquad
  \Lm_\m:=2\,\Hhat(\m\otimes\Id_r),\quad
  \mathbf{g}(\m):=\chat+\Ahat\m+\Hhat(\m\otimes\m),
  \label{eq:shifted}
\end{equation}
i.e.,\ $(\Lm_\m)_{ik}=2\sum_jH_{ijk}m_j$. If $\Hhat\in\SEP$, then
$\s\T\Hhat(\s\otimes\s)=0$, so the energy balance of \eqref{eq:shifted} for
$V_\m$ has the form \eqref{eq:VdotEP}.

\begin{theorem}[Trapping region]
\label{thm:trap}
Let $\Hhat\in\SEP$ and suppose that
$\alpha_\m:=-\lmax\bigl(\sym(\Ahat+\Lm_\m)\bigr)>0$ for some $\m$. Then the
conclusions of \cref{thm:ISS} hold for $\q-\m$, with $\alpha$ replaced by
$\alpha_\m$ and $\|\chat\|$ by $\|\mathbf{g}(\m)\|$. In particular, the ball centered at
$\m$ with radius $R_\m=(\|\mathbf{g}(\m)\|+\|\Bhat\|\bar u)/\alpha_\m$ is forward
invariant and globally attracting.
\end{theorem}
\begin{proof}
Apply the proof of \cref{thm:ISS} to $V_\m$ and \eqref{eq:shifted}.
\end{proof}

\begin{proposition}[Optimal shift by semidefinite programming]
\label{prop:sdp}
The map $\m\mapsto\lmax(\sym(\Ahat+\Lm_\m))$ is convex. The largest certified
decay rate $\alpha^\star=\max_\m\alpha_\m$ is $-t^\star$, where $t^\star$ is
the value of the semidefinite program
\begin{equation}
  \min_{t\in\R,\ \m\in\R^r}\ t\qquad\text{s.t.}\qquad
  \sym(\Ahat)+\sym(\Lm_\m)\preceq t\,\Id_r .
  \label{eq:sdp}
\end{equation}
A certificate exists if and only if $t^\star<0$.
\end{proposition}
\begin{proof}
$\m\mapsto\sym(\Ahat+\Lm_\m)$ is affine, and $\lmax$ is convex on symmetric
matrices.
\end{proof}

\noindent A search over the shift $\m$ is also used, in a sparse-identification
setting and with a different algorithmic treatment, by Kaptanoglu et
al.~\cite{Kaptanoglu2021}.

Schlegel and Noack~\cite{SchlegelNoack2015} show that, for EP systems satisfying
a non-degeneracy condition, the existence of such an $\m$ is also necessary for
a monotonically attracting trapping region. In practice $\alpha_\m$ is
maximized over $\m$, with a certificate when $\alpha^\star>0$; the radius $R_\m$ may be
reduced afterwards by a secondary minimization over the feasible set.

\subsection{Weak energy preservation: local certificates}
\label{sec:weak}

When EP holds only approximately, write $\Hhat=\Hhat_{\mathrm{EP}}+\Hhat_0$ with
$\Hhat_{\mathrm{EP}}\in\SEP$. The canonical choice is the Frobenius-orthogonal
projection (\cref{sec:soft}). Then $p(\q)=\q\T\Hhat_0(\q\otimes\q)$ and
$|p(\q)|\le\epsilon_Q\|\q\|^3$, with $\epsilon_Q:=\|\Hhat_0\|_2$, the spectral
norm of the $r\times r^2$ matrix $\Hhat_0$, since $\|\q\otimes\q\|=\|\q\|^2$.

\begin{theorem}[Local certificate under weak EP]
\label{thm:weak}
Let $\alpha>0$, $K=K(\bar u)$, $\epsilon_Q>0$ and $\alpha^2>4\epsilon_QK$.
Define
\begin{equation}
  \rho_\pm=\frac{\alpha\pm\sqrt{\alpha^2-4\epsilon_QK}}{2\epsilon_Q}.
  \label{eq:rhopm}
\end{equation}
Then $\dot V<0$ on the shell $\rho_-<\|\q\|<\rho_+$. Every ball
$\{\|\q\|\le\rho\}$ with $\rho\in[\rho_-,\rho_+)$ is forward invariant, and
every solution with $\|\q_0\|<\rho_+$ satisfies
$\limsup_{t\to\infty}\|\q(t)\|\le\rho_-$. As $\epsilon_Q\to0$,
$\rho_-\to K/\alpha$ and $\rho_+\to\infty$, recovering \cref{thm:ISS}.
\end{theorem}
\begin{proof}
From \eqref{eq:Vdot}, $\dot V\le s\,(K-\alpha s+\epsilon_Qs^2)$ with $s=\|\q\|$.
The quadratic in $s$ is negative exactly on $(\rho_-,\rho_+)$. Invariance
follows as in \cref{thm:ISS}. For $\|\q_0\|<\rho_+$ the norm cannot increase
while in the shell. On each compact sub-shell $[\rho_-+\delta,\|\q_0\|]$,
$\dot V$ is bounded away from zero, so the ball of radius $\rho_-+\delta$ is
reached in finite time, for every $\delta>0$. The limits follow from a Taylor
expansion of \eqref{eq:rhopm}.
\end{proof}

\begin{remark}[Shifted version]
\label{rem:weakshift}
The proof of \cref{thm:weak} uses only $|\s\T\Hhat(\s\otimes\s)|\le\epsilon_Q\|\s\|^3$.
Hence it applies verbatim to the shifted energy of \cref{sec:shift}, with $\alpha$
replaced by $\alpha_\m$, $\|\chat\|$ by $\|\mathbf{g}(\m)\|$, and balls centered at $\m$.
\end{remark}

\begin{remark}[Weighted energies]
All results extend to $V=\tfrac12\q\T\boldsymbol{\Xi}\q$, $\boldsymbol{\Xi}\succ0$, with EP understood
as $\q\T\boldsymbol{\Xi}\Hhat(\q\otimes\q)=0$ and $\alpha$ defined through
$\sym(\boldsymbol{\Xi}\Ahat)$. This covers non-normal $\Ahat$. However, for unknown $\boldsymbol{\Xi}$
the constraint is bilinear in $(\boldsymbol{\Xi},\Hhat)$ and no longer linear in $\Hhat$.
We leave this extension to future work.
\end{remark}

\section{Energy-preserving Bayesian operator inference}
\label{sec:bayes}

We now build the constraint $\Hhat\in\SEP$ into the posterior
\eqref{eq:jointpost}. Let $\C\in\R^{n_c\times n}$ be the matrix acting on
the joint parameter $\bth$ that applies the constraints \eqref{eq:compactC} to
the $\Hhat$-entries of all rows, with zero columns for the entries of
$\chat,\Ahat,\Bhat$. Then
\begin{equation}
  \Hhat\in\SEP\iff\C\bth=\mathbf{0},\qquad \operatorname{rank}\C=n_c .
  \label{eq:Cth}
\end{equation}
A single constraint involves up to three different rows of $\Ohat$ (e.g.\
$h_{a,bc}+h_{b,ac}+h_{c,ab}=0$). Imposing it therefore couples the $r$
row-wise regressions of \eqref{eq:rowpost}. This is the only structural
change with respect to GP-BayesOpInf. Appendix~\ref{app:toy} writes out $\C$, and the
other objects of this section, in the smallest non-trivial case $r=2$, $p=1$. Let $\Tm\in\R^{n\times(n-n_c)}$ have
columns forming a basis of $\ker\C$. We use repeatedly the following identity,
proved in Appendix~\ref{app:identity}.

\begin{lemma}
\label{lem:identity}
For $\bSig\succ0$ and $\C$ of full row rank,
\begin{equation}
  \bSig-\bSig\C\T(\C\bSig\C\T)^{-1}\C\bSig=\Tm(\Tm\T\bSig^{-1}\Tm)^{-1}\Tm\T .
  \label{eq:identity}
\end{equation}
\end{lemma}

\subsection{Hard constraint: three equivalent constructions}
\label{sec:hard}

The constraint $\C\bth=\mathbf{0}$ can be built into the posterior in three
ways, each natural from a different standpoint: restrict the prior to $\ker\C$,
so that the constraint holds before any data are seen (construction (a));
treat the constraint as a noise-free observation and condition on it
(construction (b)); or impose it on the optimization problem that defines the
maximum a posteriori estimate and read the covariance from the resulting
Hessian (construction (c)). \Cref{thm:equiv} shows that the three give one and
the same posterior.

Deterministic counterparts of two of these constructions are known: (c) is
the equality-constrained least-squares formulation of Koike and
Qian~\cite{KoikeQian2024}, and the reduced parametrization in (a) may use the
skew-symmetric representation of the EP subspace given by Gkimisis et
al.~\cite{EPrepr2025}, who exploit it through a sequential least-squares
scheme. What is new here is the probabilistic reading: the constraint defines
the \emph{support of the prior}, the three constructions give the same
\emph{posterior} and not only the same point estimate, and the posterior
covariance and exact samples are available in closed form.

\paragraph{(a) Degenerate prior on the EP subspace.} Parametrize
$\bth=\Tm\bvp$ and place a Gaussian prior
$\bvp\sim\Nor(\mathbf{0},\bSig_{\varphi,0})$ on the free coordinates. The
induced prior $\bth\sim\Nor(\mathbf{0},\Tm\bSig_{\varphi,0}\Tm\T)$ is a
degenerate Gaussian supported on $\ker\C$, so $\Prob[\Hhat\in\SEP]=1$
\emph{a priori}. The model $\zt=(\Phim\Tm)\bvp+\mathbf{e}$ is again
linear-Gaussian, with posterior
\begin{equation}
  \bvp\mid\zt\sim\Nor(\bmu_\varphi,\bSig_{\varphi}),\quad
  \bSig_{\varphi}=\bigl(\Tm\T\Phim\T\W\Phim\Tm+\bSig_{\varphi,0}^{-1}\bigr)^{-1},\quad
  \bmu_\varphi=\bSig_{\varphi}\Tm\T\Phim\T\W\zt .
  \label{eq:phipost}
\end{equation}
Here $\bSig_{\varphi,0}$ and $\bSig_\varphi$ are the prior and posterior
covariances of $\bvp$, both of size $(n-n_c)\times(n-n_c)$; the subscript $0$
marks the prior, as it does for $\bth$'s prior covariance $\bSig_0$.

\paragraph{(b) Conditioning on a pseudo-observation.} In this construction, the constraint is treated as an
observation. Introduce the \emph{pseudo-observable}
\begin{equation}
  \y:=\C\bth+\bet,\qquad \bet\sim\Nor(\mathbf{0},\epsilon^2\Id_{n_c}),
  \label{eq:pseudo}
\end{equation}
with $\bet$ independent of $\bth$ and of the state trajectory: an instrument that reports
the energy-preservation residual $\C\bth$ with accuracy $\epsilon$. Imposing EP
means conditioning on the reading $\y=\mathbf{0}$. For every $\epsilon>0$ this
is an ordinary linear-Gaussian update of $\Nor(\bmu,\bSig)$, with no
conditioning on a null event, and the pair $(\bth,\y)$ is jointly Gaussian.
Conditioning gives the Kalman measurement update
\begin{equation}
  \K_\epsilon=\bSig\C\T(\C\bSig\C\T+\epsilon^2\Id)^{-1},\qquad
  \bmu_\epsilon=\bmu-\K_\epsilon\C\bmu,\qquad
  \bSig_\epsilon=\bSig-\K_\epsilon\C\bSig ,
  \label{eq:kalman}
\end{equation}
with innovation $\mathbf{0}-\C\bmu$, the difference between the observed value
of $\y$ and the value $\C\bmu$ predicted for it before the update: the
correction is driven by the amount by which the unconstrained mean violates
EP. The derivation is given in
Appendix~\ref{app:kalman}. Since $\C\bSig\C\T\succ0$, the limit $\epsilon\to0$ exists.
We define the EP-conditioned posterior as this limit:
\begin{equation}
  \bmu_c=\bmu-\bSig\C\T(\C\bSig\C\T)^{-1}\C\bmu,\qquad
  \bSig_c=\bSig-\bSig\C\T(\C\bSig\C\T)^{-1}\C\bSig .
  \label{eq:cond}
\end{equation}

\paragraph{(c) Constrained MAP and reduced Hessian.} Consider
\begin{equation}
  \min_{\bth}\ J(\bth):=\|\Phim\bth-\zt\|_{\W}^2+\bth\T\bSig_0^{-1}\bth
  \qquad\text{s.t.}\qquad \C\bth=\mathbf{0}.
  \label{eq:cmap}
\end{equation}
The Hessian of $\tfrac12J$ is the unconstrained posterior precision
$\bSig^{-1}=\Phim\T\W\Phim+\bSig_0^{-1}$. Its \emph{reduced} Hessian is its
restriction to the feasible subspace, $\Tm\T\bSig^{-1}\Tm$, obtained by writing
$\bth=\Tm\bvp$ and differentiating twice in $\bvp$: it measures the curvature
of the objective along the directions that the constraint leaves free, and is
invertible even though $\bSig_c$ is singular.

\begin{theorem}[Equivalence]
\label{thm:equiv}
Let $\bSig^{-1}=\Phim\T\W\Phim+\bSig_0^{-1}$ be the precision of the
unconstrained posterior \eqref{eq:jointpost}, and let construction (a) use the
\emph{induced} prior $\bSig_{\varphi,0}=(\Tm\T\bSig_0^{-1}\Tm)^{-1}$. Then the
three constructions produce one and the same Gaussian $\Nor(\bmu_c,\bSig_c)$,
namely
\begin{equation}
  \bSig_c=\Tm\bigl(\Tm\T\bSig^{-1}\Tm\bigr)^{-1}\Tm\T,
  \qquad
  \bmu_c=\bSig_c\bSig^{-1}\bmu .
  \label{eq:sigmacT}
\end{equation}
Specifically:
\begin{enumerate}[label=(\roman*),leftmargin=*]
  \item \emph{(Commutation.)} Conditioning the prior $\Nor(\mathbf{0},\bSig_0)$
  on \eqref{eq:pseudo} as $\epsilon\to0$ gives
  $\Nor(\mathbf{0},\Tm\bSig_{\varphi,0}\Tm\T)$, the induced prior of
  construction (a), and the data update and the constraint update commute:
  \[
  \begin{tikzcd}[row sep=2.4em,column sep=4.2em,ampersand replacement=\&]
    \Nor(\mathbf{0},\bSig_0)
      \arrow[r,"\text{data}"]\arrow[d,"\text{constraint}"']
    \& \Nor(\bmu,\bSig)\arrow[d,"\text{constraint}"]\\
    \Nor(\mathbf{0},\Tm\bSig_{\varphi,0}\Tm\T)
      \arrow[r,"\text{data}"']
    \& \Nor(\bmu_c,\bSig_c)
  \end{tikzcd}
  \]
  \item \emph{(Degenerate prior.)} The posterior precision in
  \eqref{eq:phipost} is $\bSig_\varphi^{-1}=\Tm\T\bSig^{-1}\Tm$, and the law of
  $\Tm\bvp$ is $\Nor(\bmu_c,\bSig_c)$; equivalently
  $\bSig_c=\Tm\bSig_\varphi\Tm\T$ and $\bmu_c=\Tm\bmu_\varphi$. On $\R^n$ both
  the prior $\Nor(\mathbf{0},\Tm\bSig_{\varphi,0}\Tm\T)$ and the posterior
  $\Nor(\bmu_c,\bSig_c)$ are degenerate, of rank $n-n_c$ and supported on
  $\ker\C$; in the free coordinates $\bvp$ both are non-degenerate.
  \item \emph{(Optimization.)} The unique solution of \eqref{eq:cmap} is
  $\bmu_c$, and the inverse reduced Hessian equals $\bSig_c$:
  \[
    \Tm\bigl(\Tm\T\nabla^2(\tfrac12J)\,\Tm\bigr)^{-1}\Tm\T=\bSig_c .
  \]
  \item \emph{(Support.)} $\C\bmu_c=\mathbf{0}$ and $\C\bSig_c=\mathbf{0}$.
  Hence $\C\bth=\mathbf{0}$ almost surely: every sample of
  $\Nor(\bmu_c,\bSig_c)$ is energy-preserving.
\end{enumerate}
\end{theorem}
\begin{proof}
See Appendix~\ref{app:equiv}.
\end{proof}

\noindent Part (iii) identifies $\bmu_c$ with the solution of the
equality-constrained least-squares problem of~\cite{KoikeQian2024}, computed
there without reference to a posterior. Parts (i), (ii) and (iv) are the
statements needed for uncertainty quantification.

Since the three agree, we are free to choose the most convenient, and we use
construction (b): the EP-constrained posterior is then reached by two
successive linear-Gaussian updates of the prior $\Nor(\mathbf{0},\bSig_0)$,
\begin{equation}
\Nor(\mathbf{0},\bSig_0)\
  \xrightarrow[\ \zt=\Phim\bth+\mathbf{e}\ ]{\ \text{update 1: the data}\ }\
  \Nor(\bmu,\bSig)\
  \xrightarrow[\ \y=\C\bth+\bet=\mathbf{0}\ ]{\ \text{update 2: the constraint}\ }\
  \Nor(\bmu_c,\bSig_c)
\label{eq:twoupdates}
\end{equation}
Update~1 is the GP-BayesOpInf regression \eqref{eq:jointpost} and is the only
place where the data $(\Phim,\W,\zt)$ enter; update~2 is the pseudo-observation
update \eqref{eq:kalman}, which sees the data solely through the pair
$(\bmu,\bSig)$ produced by update~1. The constraint is therefore applied as a
post-processing step, leaving the regression of \cref{sec:background}
untouched.

\begin{remark}[Where the data enter]
\label{rem:wheredata}
Written out, \eqref{eq:sigmacT} reads
\[
  \bSig_c=\Tm\Bigl(\Tm\T\bigl(\underbrace{\Phim\T\W\Phim}_{\text{data}}
  +\underbrace{\bSig_0^{-1}}_{\text{prior}}\bigr)\Tm\Bigr)^{-1}\Tm\T ,
\]
whereas the conditioned \emph{prior} of part (i) has the same expression with
the data term removed, $\Tm\bSig_{\varphi,0}\Tm\T=\Tm(\Tm\T\bSig_0^{-1}\Tm)^{-1}\Tm\T$.
The two are distinct objects: $\bSig_{\varphi,0}$ carries no information from
$\Phim$, while $\bSig_\varphi$ does, and $\bSig_c=\Tm\bSig_\varphi\Tm\T$ is the
rank-$(n-n_c)$ image in $\R^n$ of the non-degenerate $\bSig_\varphi$.
\end{remark}

\begin{remark}[On conditioning on a null event]
$\{\C\bth=\mathbf{0}\}$ has probability zero under $\Nor(\bmu,\bSig)$, and
conditional distributions given null events are not unique in general, often referred to as the Borel--Kolmogorov paradox. \Cref{thm:equiv} fixes the construction as the
$\epsilon\to0$ limit of \eqref{eq:pseudo}. This limit coincides with the
Gaussian conditional of the jointly Gaussian pair $(\bth,\C\bth)$, with the
degenerate prior (a), and with the constrained MAP (c).
\end{remark}

\begin{remark}[Interpretation]
\label{rem:interp}
In the practically relevant form (b), enforcing EP amounts to \emph{one
Kalman measurement update} of the GP-BayesOpInf posterior, with measurement
matrix $\C$, measurement $\mathbf{0}$ and noise covariance $\epsilon^2\Id$. The GP
regression, the weights $\W_i$ and the row posteriors \eqref{eq:rowpost} are
unchanged. The correction applied to $\bmu$ is $\bSig\C\T$ times a vector, so the
posterior covariance redistributes it from the $\Hhat$-entries named by the
constraint to every entry correlated with them. Within each row the
$\Hhat$- and $\Ahat$-entries are correlated, because the linear and quadratic
features are correlated along a trajectory. The update therefore also moves and
sharpens $\Ahat$, and with it the decay rate $\alpha$ of \cref{thm:ISS}, even
though the constraint involves only $\Hhat$.
\end{remark}

\paragraph{Exact sampling.} Posterior samples are obtained without factorizing
$\bSig_c$.

\begin{proposition}[Sample correction]
\label{prop:matheron}
Let $\bth\sim\Nor(\bmu,\bSig)$ and $\bet\sim\Nor(\mathbf{0},\epsilon^2\Id)$ be
independent. Then
\begin{equation}
  \bth_\epsilon:=\bth-\K_\epsilon(\C\bth+\bet)\ \sim\ \Nor(\bmu_\epsilon,\bSig_\epsilon),
  \label{eq:matheron}
\end{equation}
and for $\epsilon=0$, with $\K_0=\bSig\C\T(\C\bSig\C\T)^{-1}$ the limit of
$\K_\epsilon$ in \eqref{eq:kalman}, $\bth_c=\bth-\K_0\C\bth\sim\Nor(\bmu_c,\bSig_c)$.
\end{proposition}
\begin{proof}
The mean is $\bmu-\K_\epsilon\C\bmu$. The covariance is
$(\Id-\K_\epsilon\C)\bSig(\Id-\K_\epsilon\C)\T+\epsilon^2\K_\epsilon\K_\epsilon\T
=\bSig-2\K_\epsilon\C\bSig+\K_\epsilon(\C\bSig\C\T+\epsilon^2\Id)\K_\epsilon\T=\bSig_\epsilon$,
using $\K_\epsilon(\C\bSig\C\T+\epsilon^2\Id)=\bSig\C\T$.
\end{proof}

Since $\bSig$ is block diagonal, $\bth\sim\Nor(\bmu,\bSig)$ is drawn row by row
exactly as in GP-BayesOpInf; the correction \eqref{eq:matheron} is a fixed
linear map (``Matheron's rule''~\cite{Wilson2020}; for $\epsilon>0$ it is the
perturbed-observation update of ensemble Kalman filtering~\cite{Evensen2009}).
The Euclidean projection $\bth\mapsto(\Id-\C\T(\C\C\T)^{-1}\C)\bth$ also yields
EP samples, but with the wrong law. It is correct only if $\bSig$ is a
multiple of the identity.

\paragraph{Closure under sequential updating.} Digital-twin applications update
the operator posterior online as new data arrive. The EP structure is preserved
by any such linear-Gaussian update.

\begin{proposition}
\label{prop:closure}
Let $\bth\sim\Nor(\bmu',\bSig')$ with $\C\bmu'=\mathbf{0}$ and $\C\bSig'=\mathbf{0}$,
and let $\y=\mathbf{G}\bth+\mathbf{e}$, $\mathbf{e}\sim\Nor(\mathbf{0},\mathbf{R})$,
$\mathbf{R}\succ0$. Then the posterior $\Nor(\bmu'',\bSig'')$ given $\y$
satisfies $\C\bmu''=\mathbf{0}$ and $\C\bSig''=\mathbf{0}$.
\end{proposition}
\begin{proof}
$\bmu''=\bmu'+\bSig'\mathbf{G}\T\mathbf{S}^{-1}(\y-\mathbf{G}\bmu')$ and
$\bSig''=\bSig'-\bSig'\mathbf{G}\T\mathbf{S}^{-1}\mathbf{G}\bSig'$, with
$\mathbf{S}=\mathbf{G}\bSig'\mathbf{G}\T+\mathbf{R}$. Multiply on the left by
$\C$ and use $\C\bSig'=\mathbf{0}$.
\end{proof}

\subsection{Soft constraint}
\label{sec:soft}

When EP is only approximately valid, take $\epsilon>0$ in
\eqref{eq:pseudo}. To interpret $\epsilon$, fix the coordinates of
$\vecop(\Hhat)$ in which the baseline prior is isotropic,
$\Nor(\mathbf{0},\gamma_{\mathrm{EP}}^{-2}\Id)$. Orthonormalize the rows of
the $\Hhat$-block of $\C$ in these coordinates (a QR factorization, which
leaves $\ker\C$ unchanged). Let $\Pm_\perp:=\C_H\T\C_H$ and
$\Pm_{\mathrm{EP}}:=\Id-\Pm_\perp$ be the orthogonal projectors onto
$\SEP^\perp$ and $\SEP$. We write $\mathbf{h}_0:=\Pm_\perp\vecop(\Hhat)$ for the
vector of compact coefficients of the non-EP part $\Hhat_0$ of
\cref{sec:weak}; $\Hhat_0$ is an $r\times r^2$ matrix and $\mathbf{h}_0$ a
vector, and we keep the two norms distinct.

\begin{proposition}[Three readings of the soft constraint]
\label{prop:soft}
The following define the same posterior:
\begin{enumerate}[label=(\roman*),leftmargin=*]
  \item the pseudo-observation \eqref{eq:pseudo} with variance $\epsilon^2$;
  \item the anisotropic prior on $\vecop(\Hhat)$ with precision
  $\gamma_{\mathrm{EP}}^2\Pm_{\mathrm{EP}}+\gamma_0^2\Pm_\perp$, where
  $\gamma_0^2=\gamma_{\mathrm{EP}}^2+\epsilon^{-2}$;
  \item the penalized MAP problem \eqref{eq:cmap} with the constraint replaced by
  the penalty $\epsilon^{-2}\|\C\bth\|^2$, together with its Hessian.
\end{enumerate}
For $\epsilon\to\infty$ one recovers GP-BayesOpInf. For $\epsilon\to0$ one
recovers \eqref{eq:cond}.
\end{proposition}
\begin{proof}
All three negative log-posteriors equal
$\tfrac12\|\Phim\bth-\zt\|^2_\W+\tfrac12\bth\T(\bSig_0^{-1}+\epsilon^{-2}\C\T\C)\bth$
up to constants, and
$\gamma_{\mathrm{EP}}^2\Id+\epsilon^{-2}\Pm_\perp=\gamma_{\mathrm{EP}}^2\Pm_{\mathrm{EP}}+(\gamma_{\mathrm{EP}}^2+\epsilon^{-2})\Pm_\perp$.
The limits follow from \eqref{eq:kalman}.
\end{proof}

\begin{remark}[Link to the local certificate]
\label{rem:softcert}
The soft posterior bounds the constant $\epsilon_Q$ of \cref{thm:weak}. Under
$\Nor(\bmu_\epsilon,\bSig_\epsilon)$ the non-EP component $\mathbf{h}_0$ is
Gaussian, and $\epsilon_Q$ is bounded by its Euclidean norm, a $1$-Lipschitz
function. Gaussian concentration~\cite[Thm.~5.6]{Boucheron2013} therefore gives
an explicit upper bound on $\epsilon_Q$, in terms of the mean and covariance of
$\mathbf{h}_0$ alone, that holds with any prescribed posterior probability.
Evaluating \eqref{eq:rhopm} at that bound certifies the local stability region
at the same confidence, since $\rho_-$ increases and $\rho_+$ decreases with
$\epsilon_Q$. The bound scales as $1/\gamma_0$, so the hyperparameter
$\gamma_0$ can be set from a physical estimate of the departure from EP.
\end{remark}

\subsection{Hyperparameter selection by marginal likelihood}
\label{sec:evidence}

We replace simulation-based screening by maximization of the marginal likelihood
of the linear-Gaussian model \eqref{eq:linmodel}. The hyperparameters are the
prior scales $\bm\gamma$ defining $\bSig_0$ --- in practice one per operator
block --- together with $\gamma_0$ for the soft constraint. The GP
hyperparameters are fitted beforehand and held fixed.

\begin{proposition}[Closed-form evidence]
\label{prop:evidence}
For $\zt=\Phim\bth+\mathbf{e}$, $\mathbf{e}\sim\Nor(\mathbf{0},\W^{-1})$,
$\bth\sim\Nor(\mathbf{0},\Pm_0^{-1})$ with $\Pm_0\succ0$, and $N=\dim\zt$,
\begin{equation}
  \log p(\zt)=-\tfrac N2\log2\pi+\tfrac12\log|\W|+\tfrac12\log|\Pm_0|
  -\tfrac12\log|\Pm|-\tfrac12\bigl(\zt\T\W\zt-\bmu\T\Pm\bmu\bigr),
  \label{eq:evidence}
\end{equation}
where $\Pm=\Phim\T\W\Phim+\Pm_0$ and $\bmu=\Pm^{-1}\Phim\T\W\zt$.
\end{proposition}
\begin{proof}
See Appendix~\ref{app:evidence}.
\end{proof}

For the soft constraint, apply \eqref{eq:evidence} with the anisotropic
precision of \cref{prop:soft}(ii). For the hard constraint, apply it in the
reduced coordinates of construction (a), with $\Phim\to\Phim\Tm$ and
$\Pm_0=\Tm\T\bSig_0^{-1}\Tm$. All quantities in \eqref{eq:evidence} are already
computed by the posterior update, and the gradient with respect to the prior
scales follows from standard matrix identities. The objective is smooth and
deterministic, so quasi-Newton methods apply. It replaces the random,
discontinuous objective of simulation-based screening. Note that the
hyperparameters should be selected \emph{under the constrained model}.
Selecting them for the unconstrained model and then conditioning is not
equivalent in general.

\section{Probabilistic stability certificates}
\label{sec:cert}

Energy preservation removes the cubic obstruction almost surely. Global
boundedness still requires the \emph{inequality} $\alpha = -\lmax\bigl(\sym(\Ahat)\bigr)>0$ (or the shifted
condition of \cref{thm:trap}). This inequality defines an open cone, not a
subspace, so conditioning on it does not preserve Gaussianity. We therefore
quantify it rather than impose it.

\subsection{Certified sets and the probability of stability}

For a margin $\delta\ge0$, define the certified sets
\begin{align}
  \mathcal S_\delta&:=\bigl\{\bth:\ \C\bth=\mathbf{0},\ \lmax(\sym\Ahat)<-\delta\bigr\},
  \label{eq:S0}\\
  \mathcal S^{\mathrm{sh}}_\delta&:=\bigl\{\bth:\ \C\bth=\mathbf{0},\
  \exists\,\m\in\R^r:\ \lmax(\sym(\Ahat+\Lm_\m))<-\delta\bigr\}\supseteq\mathcal S_\delta ,
  \label{eq:Ssh}
\end{align}
where $\Ahat,\Lm_\m$ are the operators encoded by $\bth$. By
\cref{thm:ISS,thm:trap}, every $\bth\in\mathcal S^{\mathrm{sh}}_\delta$ yields a
ROM that is globally ultimately bounded and input-to-state stable, with decay
rate at least $\delta$. We call
\begin{equation}
  p_\delta:=\Prob_{\bth\sim\Nor(\bmu_c,\bSig_c)}\bigl[\bth\in\mathcal S_\delta\bigr]
  \qquad\bigl(\text{resp.\ } p^{\mathrm{sh}}_\delta\bigr)
  \label{eq:pdelta}
\end{equation}
the \emph{posterior probability of certified stability}. Unlike a
finite-horizon simulation test, $p_\delta$ is a property of the posterior. It
holds for all initial conditions and all bounded inputs, and every factor in
it has a precise meaning. For the soft posterior $\Nor(\bmu_\epsilon,\bSig_\epsilon)$,
the set is replaced by the local certificate of \cref{thm:weak}, evaluated at a
prescribed operating radius.

\subsection{Monte Carlo estimation with exact confidence bounds}
\label{sec:mc}

Draw $\bth_1,\dots,\bth_{N_s}$ with \cref{prop:matheron}, evaluate
$\alpha_k=-\lmax(\sym\Ahat_k)$ (an $r\times r$ symmetric eigenproblem), $k=1,\dots,N_s$, and,
if $\alpha_k\le\delta$, optionally solve \eqref{eq:sdp}. Let $X$ be the number
of certified samples, and $\hat p=X/N_s$. The Clopper--Pearson
interval~\cite{ClopperPearson1934}
\begin{equation}
  \bigl[\,B^{-1}(\tfrac\beta2;X,N_s-X+1),\ B^{-1}(1-\tfrac\beta2;X+1,N_s-X)\,\bigr],
  \label{eq:cp}
\end{equation}
with $B^{-1}(\cdot;a,b)$ the Beta quantile function, has coverage at least
$1-\beta$. Since $X\sim\mathrm{Bin}(N_s,p_\delta)$, the interval
\eqref{eq:cp} is random through $X$, and it contains the true $p_\delta$ with
probability at least $1-\beta$. This holds for every $N_s$, with no normal
approximation. In particular, if all samples are certified, then $p_\delta\ge1-\varepsilon$
holds with confidence $1-\beta$ provided $N_s\ge\ln(1/\beta)/\varepsilon$
(since $(1-\varepsilon)^{N_s}\le e^{-\varepsilon N_s}$). For $\varepsilon=10^{-3}$
and $\beta=0.05$ this gives $N_s\approx3000$ eigenvalue problems of size $r$.

\subsection{The certified posterior}

The probability $p_\delta$ says how much of the posterior is certified, but
prediction requires the certified models themselves. We therefore restrict the
posterior to them.

Let $\Pi_c$ denote the probability measure $\Nor(\bmu_c,\bSig_c)$ and let
$p_\delta=\Pi_c(\mathcal S_\delta)>0$. The
\emph{certified posterior} is the conditional law
\begin{equation}
  \Pi_c(\,\cdot\mid\mathcal S_\delta)=\Pi_c(\,\cdot\cap\mathcal S_\delta)/p_\delta ,
  \label{eq:certpost}
\end{equation}
that is, $\Pi_c$ restricted to $\mathcal S_\delta$ and renormalized. Unlike
$\{\C\bth=\mathbf{0}\}$, which has probability zero and required the limiting
construction of \cref{sec:hard}, the set $\mathcal S_\delta$ has positive
probability, so conditioning on it is elementary and is achieved by rejection:
one simply discards the samples that fail the test.

\begin{proposition}[Exact sampling of the certified posterior]
\label{prop:truncated}
Samples of $\Pi_c$ accepted if and only if they lie in $\mathcal S_\delta$ are
exact samples of \eqref{eq:certpost}. The acceptance rate is $p_\delta$.
\end{proposition}
\begin{proof}
For any set $A$, $\Prob[\bth\in A\mid\bth\in\mathcal S_\delta]
=\Pi_c(A\cap\mathcal S_\delta)/p_\delta$, which is \eqref{eq:certpost}.
\end{proof}

No extra computation is involved: the draws of \cref{sec:mc} are already tested
one by one, and the accepted ones are the sample. The normalizing constant
$p_\delta$ is never needed, and is obtained as the acceptance rate. A small
$p_\delta$ makes the procedure wasteful, which is the honest cost of the method.

Restricting the posterior is not only a matter of presentation; without it the
probabilistic moments of the predicted state need not be finite. Energy preservation already rules out
blow-up in finite time, since \eqref{eq:VdotEP} bounds $\dot V$ by a quadratic
in $\|\q\|$, but samples with $\alpha\le0$ still grow exponentially, and their
contribution to $\E\|\q(t)\|$ grows without bound as $t$ increases. The
restriction removes exactly those samples.

\begin{proposition}[Moments of certified predictions]
\label{prop:moments}
Let $\delta>0$, $\bar u<\infty$ and $\bth\sim\Pi_c(\cdot\mid\mathcal S_\delta)$.
Then $\sup_{t\ge0}\E\|\q(t)\|^\ell<\infty$ for every $\ell\ge1$.
\end{proposition}
\begin{proof}
By \cref{thm:ISS} and $\alpha>\delta$,
$\|\q(t)\|\le\|\q_0\|+(\|\chat\|+\|\Bhat\|\bar u)/\delta$ for all $t$. The
right-hand side does not depend on $t$ and is an affine function of norms of
Gaussian vectors, so it has finite moments of every order under $\Pi_c$.
Conditioning on an event of probability $p_\delta>0$ multiplies them by at most
$1/p_\delta$.
\end{proof}

Uniformity in $t$ is what makes the supremum finite, and it requires $\delta>0$
strictly: the ultimate bound is proportional to $1/\delta$, and models with
arbitrarily slow decay would not have integrable bounds.
\Cref{prop:moments} is what justifies reporting posterior predictive means and
variances over the whole time horizon.

\begin{remark}[Comparison with simulation-based screening]
Simulation-based screening rejects \emph{hyperparameters} on the basis of
trajectory amplitudes over a finite horizon, for one initial condition, using
$\mathcal O(10)$ ODE solves. \Cref{prop:truncated} rejects \emph{samples} on the
basis of a global certificate, using $\mathcal O(10^3$--$10^4)$ eigenvalue
problems of size $r$. It returns a well-defined distribution together with its
normalizing probability $p_\delta$, which is reported rather than hidden.
\end{remark}

\section{Conservative systems: constraining the linear part}
\label{sec:conservative}

The certificates of \cref{sec:cert} ask for a decay rate, and a large class of
problems has none to offer. A semidiscrete system is called \emph{conservative} when
its right-hand side neither creates nor destroys the energy
$E=\tfrac12\|\qq\|^2$: with $\bu=\mathbf{0}$ this means
\begin{equation}
  \qq\T\Hm_F(\qq\otimes\qq)=0\quad\text{for all }\qq,
  \qquad \sym(\mathbf{A}_F)=\mathbf{0},
  \label{eq:consFOM}
\end{equation}
so that $\mathrm dE/\mathrm dt=0$ along every trajectory. Centered
discretizations of transport and dispersion are of this kind, as are the
spatial discretizations of the Korteweg--de Vries, Euler and shallow-water
equations away from forcing and dissipation; \cref{sec:prob2} uses one.
These systems make the structure imposed here especially valuable and the
certificates of \cref{sec:cert} especially uninformative: nothing damps an
error once it is made, and \cref{thm:ISS,thm:trap} certify an attracting ball
that the equation does not possess. The purpose of this section is to say what replaces the certificate
there.

\subsection{Skew-symmetric linear part}

The first condition in \eqref{eq:consFOM} is energy preservation of the
quadratic term and is inherited by Galerkin projection
(\cref{prop:galerkin}). The second is inherited in the same way: if
$\sym(\mathbf{A}_F)=\mathbf{0}$, then $\V\T\mathbf{A}_F\V$ is skew-symmetric for
every $\V$ with orthonormal columns. It is therefore natural to ask of the
inferred linear operator what the projected one satisfies. Write
\begin{equation}
  \SSK:=\{\Ahat\in\R^{r\times r}\ :\ \sym(\Ahat)=\mathbf{0}\}.
  \label{eq:skewset}
\end{equation}
Since $\sym(\Ahat)=\tfrac12(\Ahat+\Ahat\T)$, the condition is exactly
$\Ahat\T=-\Ahat$: the members of $\SSK$ are the skew-symmetric matrices, whose
strictly upper triangle is free and determines the rest. Hence $\SSK$ is a
linear subspace of dimension $r(r-1)/2$, defined by the $r(r+1)/2$ independent
equalities
\begin{equation}
  \hat A_{ii}=0\quad(1\le i\le r),
  \qquad
  \hat A_{ij}+\hat A_{ji}=0\quad(i<j),
  \label{eq:skewC}
\end{equation}
one for each entry on or above the diagonal of the symmetric matrix
$\sym(\Ahat)$.
Asking for both structures at once gives a bound that holds for every operator
in the family, with no inequality to check.

\begin{theorem}[Linear growth]
\label{thm:skew}
Let $\Hhat\in\SEP$ and $\Ahat\in\SSK$. Then, with $K(\bar u)$ as in
\eqref{eq:alphaK},
\begin{equation}
  \|\q(t)\|\le\|\q_0\|+K(\bar u)\,t\qquad\text{for all }t\ge0 ,
  \label{eq:lingrowth}
\end{equation}
whatever $\chat$, $\Bhat$, $\q_0$ and whatever the spectrum of $\Ahat$. If in
addition $\chat=\mathbf{0}$ and $\bu\equiv\mathbf{0}$, then the reduced energy is
conserved exactly: $\|\q(t)\|=\|\q_0\|$ for all $t$.
\end{theorem}
\begin{proof}
Both quadratic forms in \eqref{eq:Vdot} vanish identically:
$\q\T\Hhat(\q\otimes\q)=0$ since $\Hhat\in\SEP$, and
$\q\T\Ahat\q=\q\T\sym(\Ahat)\q=0$ since $\Ahat\in\SSK$. Hence
\begin{equation}
  \dot V=\q\T\chat+\q\T\Bhat\bu\le K(\bar u)\,\|\q\| ,
  \label{eq:VdotCons}
\end{equation}
which is $0$ when $\chat=\mathbf{0}$ and $\bu\equiv\mathbf{0}$, giving the second
claim. For the first, $\|\q\|=\sqrt{2V}$ is not differentiable where
$\q=\mathbf{0}$, so fix $\varepsilon>0$ and use the smoothed norm
$W_\varepsilon:=\sqrt{2V+\varepsilon^2}=\sqrt{\|\q\|^2+\varepsilon^2}$, whose
radicand never vanishes. It is differentiable in $t$, and by
\eqref{eq:VdotCons},
\begin{equation}
  \frac{\mathrm dW_\varepsilon}{\mathrm dt}
  =\frac{\dot V}{\sqrt{\|\q\|^2+\varepsilon^2}}
  \le K(\bar u)\,\frac{\|\q\|}{\sqrt{\|\q\|^2+\varepsilon^2}}
  \le K(\bar u) ,
  \label{eq:Weps}
\end{equation}
where the last step is because $\|\q\|\le\sqrt{\|\q\|^2+\varepsilon^2}$. Integrating
\eqref{eq:Weps} from $0$ to $t$ gives
\begin{equation*}
  \|\q(t)\|\le W_\varepsilon(t)
  =W_\varepsilon(0)+\int_0^t\frac{\mathrm dW_\varepsilon}{\mathrm ds}(s)\,\mathrm ds
  \le\sqrt{\|\q_0\|^2+\varepsilon^2}+K(\bar u)\,t ,
\end{equation*}
and letting $\varepsilon\to0$ gives \eqref{eq:lingrowth}.
\end{proof}

\Cref{thm:skew} is weaker than \cref{thm:GES,thm:ISS}: it gives no attracting
set and no decay, only that nothing blows up in finite time and that the energy
moves no faster than the constant term allows. For a conservative problem that
is the right conclusion: the equation conserves energy, so it neither grows nor
decays, and a reduced model that decayed would misrepresent the physics. Since
it asks for no condition on the operators, the bound holds for every draw of
the posterior below rather than for a certified subset, so no probability has
to be estimated and \cref{sec:cert} is not used.

\begin{remark}[Why the certificates say nothing here]
\label{rem:certvacuous}
Under $\SSK$ the certified sets \eqref{eq:S0} are empty by construction:
$\lmax(\sym\Ahat)=0$ exactly, so the strict inequality $\lmax(\sym\Ahat)<-\delta$
fails for every $\delta\ge0$ and $p_\delta=0$. This is a property of the model
class, not an outcome of the inference, and it is the reason why the certificates
are reported in \cref{sec:results2} for the energy-preserving posterior alone.
\end{remark}

\subsection{The posterior}
\label{sec:conspost}

Imposing \eqref{eq:skewC} requires no new machinery: it adds rows to the
constraint matrix of \cref{sec:bayes}. Let $\C_A\in\R^{n_a\times n}$, with
$n_a=r(r+1)/2$, apply \eqref{eq:skewC} to the $\Ahat$-entries of all rows of
$\Ohat$, with zero columns elsewhere, so that, in the notation of
\eqref{eq:Cth},
\begin{equation}
  \Ahat\in\SSK\iff\C_A\bth=\mathbf{0},
  \qquad
  \C_{HA}:=\begin{bmatrix}\C\\[2pt]\C_A\end{bmatrix},
  \qquad
  \Hhat\in\SEP\ \text{and}\ \Ahat\in\SSK\iff\C_{HA}\bth=\mathbf{0}.
  \label{eq:CHAth}
\end{equation}
The two blocks act on disjoint coordinates of $\bth$, so $\C_A$ has full row
rank $n_a$, $\operatorname{rank}\C_{HA}=n_c+n_a$, and the constrained subspace
has dimension $n-n_c-n_a$.

Everything in \cref{sec:bayes} is stated for a matrix of full row rank and for
its kernel, and therefore holds verbatim with $\C$ replaced by $\C_{HA}$. The
posterior is again the Kalman update \eqref{eq:cond} of the GP-BayesOpInf
posterior, now with innovation $\mathbf{0}-\C_{HA}\bmu$; it is again degenerate
with support $\ker\C_{HA}$, so both structures hold almost surely rather than
approximately; its three readings in \cref{thm:equiv} are unchanged; and exact
samples are again produced by the single linear correction of
\cref{prop:matheron},
\begin{equation}
  \bth_c=\bth-\K_{HA}\C_{HA}\bth,
  \qquad
  \K_{HA}=\bSig\C_{HA}\T(\C_{HA}\bSig\C_{HA}\T)^{-1} .
  \label{eq:mathHA}
\end{equation}
The soft version of \cref{sec:soft} and the evidence of \cref{prop:evidence}
carry over in the same way, the latter with $\Tm$ replaced by a basis of
$\ker\C_{HA}$. The algorithms of \cref{sec:alg} are unchanged except for the
row count of $\C$.

\begin{remark}[The sparsity pattern survives]
\label{rem:sparseHA}
The structure used in \cref{rem:attr-count} and \cref{sec:soft} is preserved.
Each row of $\C_A$ has one or two nonzero entries, each entry of $\Ahat$ appears
in exactly one of them, and no coordinate of $\bth$ is touched by both blocks.
Hence $\C_{HA}$ still has one nonzero per column, its rows are still mutually
orthogonal, and
$\C_{HA}\C_{HA}\T=\diag(\nu_1,\dots,\nu_{n_c+n_a})$ with $\nu_\ell\in\{1,2,3\}$,
so orthonormalizing the rows remains a rescaling by $\nu_\ell^{-1/2}$ rather
than a factorization. The cost of the second structure is one block of extra
rows and nothing else.
\end{remark}

\section{Algorithms and computational cost}
\label{sec:alg}

Everything up to the row posteriors \eqref{eq:rowpost} is
GP-BayesOpInf~\cite{McQuarrie2025}: the POD compression, the Gaussian process
fits that return the state estimates, the time-derivative estimates and their
weight matrices, the data matrix $\Dt$, and the row-wise Gaussian posteriors
themselves. We take all of it unchanged and do not restate it here.

\begin{figure}[htbp]
\centering
\begin{tikzpicture}[
  node distance=6mm and 22mm,
  blk/.style={rounded corners=2pt, draw=black!55, align=center, inner sep=4pt,
              font=\small, text width=52mm},
  data/.style={blk, fill=violet!10},
  red/.style={blk, fill=cyan!10},
  gp/.style={blk, fill=green!12},
  inf/.style={blk, fill=black!6},
  new/.style={blk, fill=orange!18, draw=orange!75!black, line width=1pt},
  lbl/.style={font=\scriptsize, fill=black!5, rounded corners=1pt, inner sep=2pt},
  newlbl/.style={lbl, fill=orange!18, draw=orange!75!black},
  ar/.style={-{Stealth[length=5pt]}, draw=black!65},
]
\node[data] (snap) {noisy, sparse snapshots\\[1pt]
  $\qq_1,\dots,\qq_m\in\R^{N}$};
\node[red, below=10mm of snap] (red) {POD coefficient data\\[1pt]
  $\y_1,\dots,\y_r\in\R^{m}$};
\node[gp, below=10mm of red] (gp) {GP state and derivative estimates\\[1pt]
  weights $\W_i$, data matrix $\Dt$};
\node[inf, below=10mm of gp] (post) {GP-BayesOpInf posterior\\[1pt]
  $\ohat_i\sim\Nor(\bmu_i,\bSig_i)$, \eqref{eq:rowpost}};

\draw[ar] (snap) -- node[lbl, right] {compression} (red);
\draw[ar] (red) -- node[lbl, right] {Gaussian process regression} (gp);
\draw[ar] (gp) -- node[newlbl, right] {prior scales by evidence \eqref{eq:evidence}} (post);

\node[new, right=of post] (ep) {energy-preserving posterior\\[1pt]
  $\bmu_c=\bmu-\K\C\bmu$, $\bSig_c=\bSig-\K\C\bSig$\\[1pt]
  one Kalman update, $\C$ from Algorithm~\ref{alg:C}};
\node[new, above=10mm of ep] (cert) {stability certificate\\[1pt]
  $\alpha_k=-\lmax(\sym\Ahat_k)$, shift \eqref{eq:sdp}\\[1pt]
  $\hat p_\delta$ with interval \eqref{eq:cp}};
\node[inf, above=10mm of cert] (solve) {sample certified operators,\\[1pt]
  solve the reduced model};
\node[data, above=10mm of solve] (pred) {predictions with uncertainty\\[1pt]
  $\qq(t)\approx\V\q(t;\bth)+\bar\qq$};

\draw[ar] (post) -- (ep);
\draw[ar] (ep) -- node[lbl, left] {exact draws, \cref{prop:matheron}} (cert);
\draw[ar] (cert) -- (solve);
\draw[ar] (solve) -- node[lbl, right] {decompression} (pred);
\end{tikzpicture}
\caption{The method in one picture. The pipeline is
GP-BayesOpInf~\cite{McQuarrie2025}; the shaded items are this work. Energy
preservation enters the Gaussian posterior as a single Kalman update, the prior
scales come from the marginal likelihood rather than from a search over
simulated trajectories, and the certificate replaces the simulation test. None
of the three leaves closed form.}
\label{fig:pipeline}
\end{figure}
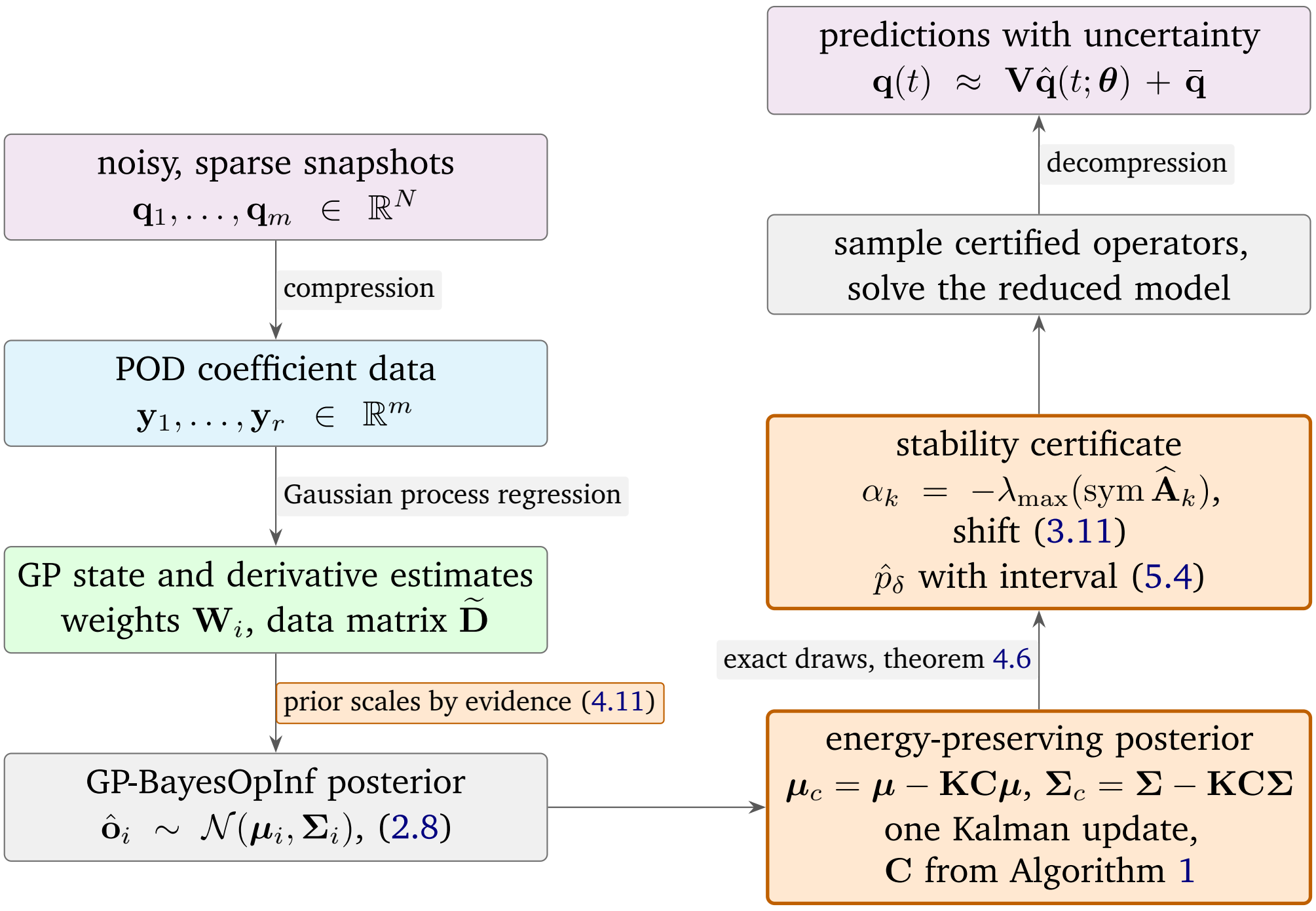

What this work adds enters at three points: i) Energy preservation is linear in
the operator coefficients, so conditioning the posterior on it is a single
Kalman update; ii) The prior scales are chosen by maximizing the marginal
likelihood instead of by the simulation search of~\cite{McQuarrie2025}; And iii)
stability is reported by the certificate of \cref{sec:cert}, evaluated draw by
draw, rather than by integrating a few samples over a finite window.
\Cref{fig:pipeline} places all three against the inherited pipeline, and the
algorithms below give only what is new.

\paragraph{The constraint matrix.} \Cref{alg:C} writes the $n_c=\binom{r+2}3$
rows of $\C$ directly, one per cubic monomial, with at most three nonzeros
each. It is built once, from $r$ alone, and never from data. We write
$\mathrm{idx}(i,\cdot)$ for the position in $\bth$ of a coefficient of row $i$,
in the compact ordering of $\Hhat$.

\begin{algorithm}[htbp]
\caption{Assemble the energy-preservation constraint matrix $\C$}
\label{alg:C}
\begin{algorithmic}[1]
\Require reduced dimension $r$, index map $\mathrm{idx}$
\Ensure sparse $\C\in\R^{n_c\times n}$, $n_c=\binom{r+2}3$
\State $\ell\gets0$
\For{$a=1,\dots,r$}\ \Comment{monomial $\hat q_a^3$}
  \State $\ell\gets\ell+1$;\quad $C_{\ell,\mathrm{idx}(a,aa)}\gets1$
\EndFor
\For{$1\le a<c\le r$}
  \State $\ell\gets\ell+1$;\quad $C_{\ell,\mathrm{idx}(a,ac)}\gets1$;\quad $C_{\ell,\mathrm{idx}(c,aa)}\gets1$ \Comment{$\hat q_a^2\hat q_c$}
  \State $\ell\gets\ell+1$;\quad $C_{\ell,\mathrm{idx}(a,cc)}\gets1$;\quad $C_{\ell,\mathrm{idx}(c,ac)}\gets1$ \Comment{$\hat q_a\hat q_c^2$}
\EndFor
\For{$1\le a<b<c\le r$}\ \Comment{$\hat q_a\hat q_b\hat q_c$}
  \State $\ell\gets\ell+1$;\quad $C_{\ell,\mathrm{idx}(a,bc)}\gets1$;\ $C_{\ell,\mathrm{idx}(b,ac)}\gets1$;\ $C_{\ell,\mathrm{idx}(c,ab)}\gets1$
\EndFor
\end{algorithmic}
\end{algorithm}

\paragraph{The update.} \Cref{alg:post} is the whole of \cref{thm:equiv} in
five lines: a gain $\K$, a corrected mean, and, if the covariance is wanted, a
corrected covariance. Setting $\epsilon>0$ gives the soft variant of
\cref{sec:soft}; $\epsilon=0$ is the hard constraint. Nothing here is iterative.

\begin{algorithm}[htbp]
\caption{Energy-preserving posterior (hard: $\epsilon=0$; soft: $\epsilon>0$)}
\label{alg:post}
\begin{algorithmic}[1]
\Require row posteriors $(\bmu_i,\bSig_i)_{i=1}^r$; $\C$; $\epsilon\ge0$
\Ensure gain $\K$, mean $\bmu_c$ and, if needed, covariance $\bSig_c$
  \Statex\hspace{\algorithmicindent}(written $\bmu_\epsilon,\bSig_\epsilon$ when $\epsilon>0$, as in \cref{sec:soft})
\State $\bmu\gets[\bmu_1;\dots;\bmu_r]$;\quad $\bSig\gets\blkdiag(\bSig_1,\dots,\bSig_r)$ \Comment{never formed densely}
\State $\mathbf{G}\gets\bSig\C\T$ \Comment{block-wise; $\C$ has $\le3$ nonzeros per row}
\State $\mathbf{S}\gets\C\mathbf{G}+\epsilon^2\Id_{n_c}$;\quad Cholesky $\mathbf{S}=\mathbf{L}_S\mathbf{L}_S\T$
\State $\K\gets\mathbf{G}\,\mathbf{S}^{-1}$ \Comment{two triangular solves}
\State $\bmu_c\gets\bmu-\K(\C\bmu)$ \Comment{corrected mean}
\State $\bSig_c\gets\bSig-\K\mathbf{G}\T$ \Comment{only if the covariance itself is needed}
\end{algorithmic}
\end{algorithm}

\paragraph{Drawing and certifying.} \Cref{alg:sample} draws from the
unconstrained posterior and corrects each draw by the same gain, which is exact
by \cref{prop:matheron}, so no constrained factorization is ever formed. Each
draw is then certified, or not, by an eigenvalue problem of size $r$ and, when
the plain test fails, one small semidefinite program.

\begin{algorithm}[htbp]
\caption{Sampling, certification and reporting}
\label{alg:sample}
\begin{algorithmic}[1]
\Require $(\bmu_i,\bSig_i)$ with Cholesky factors $\mathbf{L}_i$; $\C$, $\K$, $\epsilon$; margin $\delta\ge0$; input bound $\bar u$; $N_s$; confidence $1-\beta$
\Ensure $\hat p_\delta$ and interval \eqref{eq:cp}; certified draws $\mathcal A$; radii $\{R_k\}$
\State $X\gets0$;\quad $\mathcal A\gets\emptyset$
\For{$k=1,\dots,N_s$}
  \State draw $\ohat_i\gets\bmu_i+\mathbf{L}_i\bm\xi_i$, $\bm\xi_i\sim\Nor(\mathbf{0},\Id_d)$, and set $\bth\gets[\ohat_1;\dots;\ohat_r]$
  \State $\bth_k\gets\bth-\K(\C\bth+\bet)$, $\bet\sim\Nor(\mathbf{0},\epsilon^2\Id)$ \Comment{\cref{prop:matheron}}
  \State extract $\chat_k,\Ahat_k,\Hhat_k,\Bhat_k$;\quad $\alpha_k\gets-\lmax(\sym\Ahat_k)$;\quad $\mathbf{g}_k\gets\chat_k$
  \If{$\alpha_k\le\delta$}
    \State solve \eqref{eq:sdp} for $(\m_k,t_k)$;\quad $\alpha_k\gets-t_k$;\quad $\mathbf{g}_k\gets\chat_k+\Ahat_k\m_k+\Hhat_k(\m_k\otimes\m_k)$
  \EndIf
  \State certified iff $\alpha_k>\delta$, and, if $\epsilon>0$, also $\alpha_k^2>4\epsilon_{Q,k}(\|\mathbf{g}_k\|+\|\Bhat_k\|\bar u)$ with $\epsilon_{Q,k}=\|\Pm_\perp\vecop(\Hhat_k)\|$
  \If{certified}
    \State $X\gets X+1$;\quad $\mathcal A\gets\mathcal A\cup\{\bth_k\}$;\quad $R_k\gets(\|\mathbf{g}_k\|+\|\Bhat_k\|\bar u)/\alpha_k$ \Comment{or $\rho_-$ of \eqref{eq:rhopm}}
  \EndIf
\EndFor
\State $\hat p_\delta\gets X/N_s$ and its interval \eqref{eq:cp}; integrate the ROM for $\bth\in\mathcal A$ and report quantiles and moments (\cref{prop:moments})
\end{algorithmic}
\end{algorithm}

\paragraph{Prior scales.} GP-BayesOpInf searches over candidate scales and
keeps the candidate that reproduces the training data best among those whose
sampled ROMs stay bounded over a finite window, so every candidate costs a set
of trajectory solves. We maximize the marginal likelihood \eqref{eq:evidence}
instead, which costs little: it is evaluated in the reduced coordinates
$\Phim\Tm$ with prior $\Pm_0=\Tm\T\bSig_0(\bm\gamma)^{-1}\Tm$ under the hard
constraint, and with the anisotropic prior of \cref{prop:soft}(ii) under the
soft one, and maximized in $\log\bm\gamma$ by L-BFGS on a smooth, deterministic
objective. The GP kernel hyperparameters are fitted upstream by their own
marginal likelihood, as in~\cite{McQuarrie2025}, and held fixed throughout.

\paragraph{Cost.} Let $d=1+r+r(r+1)/2+p$, $n=rd$ and $n_c=\binom{r+2}3$. Steps
1--2 of \cref{alg:post} cost $\mathcal O(n_cd)$, because each column of $\C\T$
touches at most three blocks of size $d$, and two constraints interact only if
they share a row index. The Cholesky factorization of $\mathbf{S}$ dominates at
$\frac13n_c^3$ flops: about $1.2\times10^9$ for $r=20$ and $5\times10^{11}$ for
$r=40$, that is, well under a minute on a workstation, and it is paid once.
Each draw in \cref{alg:sample} costs $\mathcal O(rd^2)$ for the row draws,
$\mathcal O(n_c^2)$ for the correction and $\mathcal O(r^3)$ for the eigenvalue
problem; the semidefinite program has $r+1$ scalar variables and one $r\times r$
inequality. For substantially larger $r$, the reduced parametrization of
\cref{thm:equiv}(i), of dimension $n-n_c$, can be combined with solvers that
exploit the Kronecker structure $\Phim=\Id_r\otimes\Dt$.

\section{Numerical experiments}
\label{sec:numerics}

We study a forced viscous Burgers’ equation and an unforced Korteweg--de Vries
equation, both in one spatial dimension on a periodic domain, and a
two-dimensional Burgers’ equation. Energy preservation
is not implied by convection alone: it must hold for the spatially discrete
quadratic operator. We therefore use energy-preserving semidiscretizations; by
\cref{prop:galerkin}, their intrusive Galerkin ROMs inherit this property when
projected in the corresponding discrete $L^2$ inner product. The unconstrained
GP-BayesOpInf posterior of~\cite{McQuarrie2025} serves as a baseline for
assessing the effect of imposing this structure on the operator posterior.

The first problem is developed in full: the energy identity, the choice of
discretization and the reading of every reported quantity are given there once.
The second and third share the convective term, its discretization and the
observation protocol, so only what differs is stated.

\paragraph{Observation model and reported quantities.}

The observation model follows~\cite{McQuarrie2025}. For each of $L=5$ parameter
values we integrate the full-order model, observe $m$ snapshots at random times
in the training window, and add relative Gaussian noise of level $\xi$; a
single POD basis of dimension $r$ is built from all noisy snapshots, centered on
their mean $\bar\qq$, and a
Gaussian process is fitted to each of the $Lr$ reduced coordinate series. Data
are withheld in two ways, and both are used only to measure prediction error: no
snapshot is taken from the second half of the time domain, and one further
parameter value $(a^\star,b^\star)$, not among the $L$ used for training, is reserved
entirely. Nominal values are
$r=8$, $m=30$, $\xi=3\%$ for problem~1, $r=10$, $m=120$, $\xi=3\%$ for problem~2,
and $r=6$, $m=30$, $\xi=3\%$ for problem~3. Each reported configuration uses one fixed random seed, stated with
the code. Prior scales are selected by maximizing the evidence of
\cref{prop:evidence}; no trajectory is integrated to choose a hyperparameter.
They are selected separately for each posterior: under the constrained model for
ours and under the unconstrained model for the baseline, since reusing the
constrained scales would handicap the latter.

Three quantities are reported, in each case for the constrained and the
unconstrained posterior.
\begin{enumerate}[label=(\roman*),leftmargin=*,nosep]
  \item \emph{Constraint violation} $\max_k\|\C\bth_k\|$ over posterior draws,
  which \cref{thm:equiv}(iv) predicts to be at machine precision under the
  constraint, together with the relative energy-preservation defect of the
  posterior mean.
  \item \emph{Accuracy}: relative state error against the full-order solution
  on the training and prediction windows, for the training trajectories and for
  $(a^\star,b^\star)$, with the projection error of the full-order solution as the
  attainable floor.
  \item \emph{Stability}: the certified probability $\hat p_\delta$ with its
  Clopper--Pearson interval \eqref{eq:cp} from $N_s=3000$ draws, and the
  posterior distributions of $\alpha$ and $R$ from \eqref{eq:alphaK}.
\end{enumerate}

\subsection{Problem~1: forced viscous Burgers}
\label{sec:prob1}

On the periodic domain $x\in[0,1[$,
\begin{equation}
  \frac{\partial w}{\partial t}+w\frac{\partial w}{\partial x}
  =\nu\frac{\partial^2w}{\partial x^2}+\phi_1(x)\,u_1(t)+\phi_2(x)\,u_2(t),
  \qquad w(x,0)=\tfrac15\sin2\pi x,
  \label{eq:burgers1d}
\end{equation}
with actuator profiles and input
\begin{equation}
  \phi_1(x)=\sin^3(2\pi x),
  \qquad
  \phi_2(x)=\cos^3(2\pi x),
  \qquad
  \bu(t;a,b)=\begin{pmatrix}a\sin2\pi t\\ b\sin4\pi t\end{pmatrix}.
  \label{eq:actuators}
\end{equation}
The profiles are smooth, zero-mean, and one-periodic. They are quarter-period
shifts of one another and are orthogonal in space,
$\int_0^1\phi_1\phi_2\,\mathrm dx=0$.
The cubic profiles contain both the first and third spatial harmonics, unlike
single-mode sinusoidal actuators. Since the forcing has zero spatial mean and
the periodic convection and diffusion terms also have zero mean, the zero mean
of the initial condition is preserved. The amplitude of the initial condition
is of the same order as the response the actuators sustain, so that the training
window is not dominated by a decaying transient. The initial condition is the
same for every trajectory; only $(a,b)$ varies.

\paragraph{Energy preservation and full-order discretization.}
Let $E(t)=\tfrac12\int_0^1w^2\,\mathrm dx$ denote the energy, the continuous
counterpart of the reduced Lyapunov function~\eqref{eq:V}. With
$f=\phi_1u_1+\phi_2u_2$, differentiation and integration by parts give
\begin{equation}
  \frac{\mathrm dE}{\mathrm dt}
  =-\int_0^1w^2w_x\,\mathrm dx
    +\nu\int_0^1w\,w_{xx}\,\mathrm dx
    +\int_0^1w f\,\mathrm dx
  =-\nu\int_0^1w_x^2\,\mathrm dx+\int_0^1w f\,\mathrm dx.
  \label{eq:burgersenergy}
\end{equation}
The convective contribution vanishes because
$\int_0^1w^2w_x\,\mathrm dx=[w^3/3]_0^1=0$ under periodic boundary
conditions, as the two limits are the
same point; likewise, periodicity removes the boundary term when the viscous
term is integrated by parts. Thus convection preserves energy, viscosity dissipates
it, and forcing can inject or remove it. Energy preservation here refers to
the quadratic convective term, as in \cref{def:ep}, not to the full forced,
viscous equation.

The data of this experiment come from a simulation rather than from
measurement, so the discretization is part of the experimental setup and is
reported here for reproducibility. Trajectories collected from the physical
system would satisfy \eqref{eq:burgersenergy} by construction; in a simulation
the property has to be built into the discrete operator, because a continuum
energy identity does not automatically survive spatial discretization.
Let $x_i=i\Delta x$, $i=0,\ldots,N-1$, with indices interpreted
periodically, and write $w_i(t)=w(x_i,t)$ and
$\qq=(w_0,\ldots,w_{N-1})\T$. Centered differences applied to $ww_x$ are not
energy preserving in general. Since $\partial_x(w^2)=2ww_x$, the convective
term can be split as $\lambda\,\tfrac12\partial_x(w^2)+(1-\lambda)\,ww_x$ for
any $\lambda$, and discretizing the two parts by centered differences gives a
one-parameter family of second-order approximations. All members agree in the
continuum, but only $\lambda=\tfrac23$ preserves the discrete energy
(Appendix~\ref{app:lambda}), and it gives
\begin{equation}
  \bigl[\Hm_F(\qq\otimes\qq)\bigr]_i
  =-\frac13\,\frac{w_{i+1}^2-w_{i-1}^2}{2\Delta x}
   -\frac13\,w_i\,\frac{w_{i+1}-w_{i-1}}{2\Delta x}.
  \label{eq:skewsym}
\end{equation}
We verify that $\qq\T\Hm_F(\qq\otimes\qq)$ vanishes to machine precision on
random states before any inference is performed; otherwise every subsequent
statement about energy preservation would be a statement about the
discretization rather than about the method. The viscous term makes
$\sym(\mathbf{A}_F)\neq\mathbf{0}$, so the constraint of \cref{sec:conservative}
does \emph{not} apply here, since imposing it would delete the viscosity.

We take $\nu=10^{-2}$, a Reynolds number of $10^{2}$. The equation is
discretized on a uniform periodic grid with spacing $\Delta x=2^{-10}$, giving a
full-order state dimension of $N=1024$, and integrated over $t\in[0,2]$ with an
adaptive explicit Runge--Kutta scheme; the training window is $[0,1]$.

\subsubsection{Results for problem~1}
\label{sec:results1}

\begin{figure}[htbp]
\centering
\includegraphics[width=\textwidth]{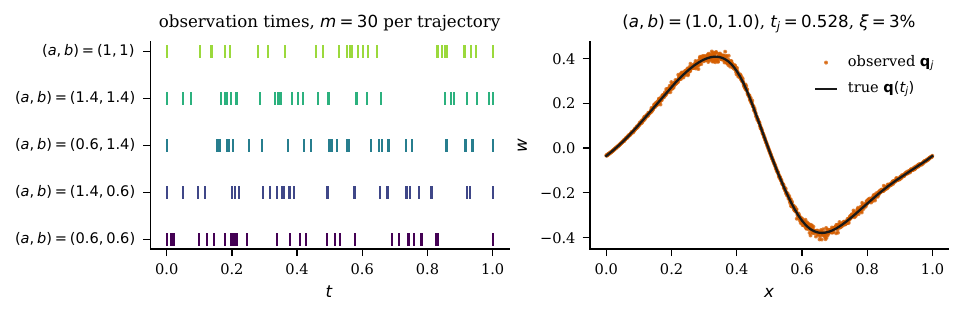}
\caption{What the method is given for problem~1. Left: the $m=30$ observation
times of each training trajectory, drawn independently and uniformly in $[0,1]$
with the endpoints included. Right: one observed snapshot against the solution it
was taken from, at relative noise level $\xi=3\%$. The reference solution of the
right panel, the second half of the time domain and the held-out
$(a^\star,b^\star)$ are never seen by the inference.}
\label{fig:prob1data}
\end{figure}

\Cref{fig:prob1data} shows the data. The reduced dimension is $r=8$, which retains $0.99635$ of the centered snapshot energy,
so $d=47$, $n=rd=376$ and $n_c=\binom{10}{3}=120$. The constraint matrix has
full rank $120$, $\ker\C$ has dimension $256$, and $\|\C\Tm\|=3.3\times10^{-16}$.

\paragraph{Exactness.}
As \cref{thm:equiv}(iv) predicts, every draw of the constrained posterior
satisfies \eqref{eq:EPindex} to machine precision: $\|\C\bth_k\|\le
3.7\times10^{-14}$ over $200$ draws, and $\|\C\bmu\|=3.0\times10^{-14}$ at the
mean. The unconstrained posterior violates it by $\|\C\bmu\|=25.8$, with a median
of $28.8$ over draws, so the gap is structural rather than numerical. Constructions~(a)
and~(b) of \cref{thm:equiv} agree elementwise to $5.7\times10^{-11}$.

\paragraph{Prior scales.}
Maximizing \eqref{eq:evidence} under each model gives scales of the same order,
the largest difference being on $\Hhat$, which the constrained model shrinks
nearly twice as hard. The unconstrained evidence is the larger of the two,
$\log p(\zt)=5790.6$ against $5226.7$, as it must be, since removing the constraint can only improve the fit
to the observed derivatives; the accuracy reported below shows that this does not
translate into predictive skill.

\begin{figure}[htbp]
\centering
\includegraphics[width=\textwidth]{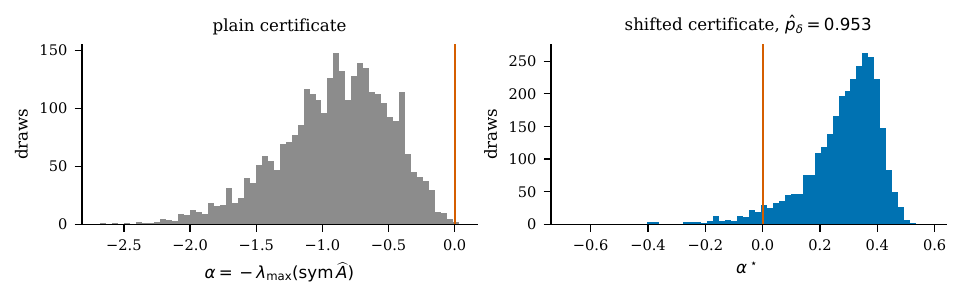}
\caption{Problem~1: the two certificates over $N_s=3000$ draws of the
constrained posterior. The plain certificate of \eqref{eq:alphaK} is negative on
all but two draws; the shifted certificate \eqref{eq:sdp} is positive on
$95.33\%$ of them. The vertical line marks the threshold $\delta=0$.}
\label{fig:prob1cert}
\end{figure}

\paragraph{Stability.}
At the constrained posterior mean the linear part is not dissipative,
$\alpha=-0.771$, and \eqref{eq:alphaK} certifies only two of the $3000$ draws.
The shifted certificate \eqref{eq:sdp} succeeds: at the posterior mean
$\alpha^\star=+0.364$ with $\|\m^\star\|=3.90$, and over the draws
\begin{equation*}
  \hat p_\delta = 0.9533,
  \qquad
  \text{Clopper--Pearson } 95\%:\ [0.9452,\ 0.9606],
\end{equation*}
with $\alpha^\star$ ranging over $[-0.676,\,0.578]$ and median $0.309$;
\cref{fig:prob1cert} shows both distributions. The
certified radius has median $R=126$ and ninetieth percentile $266$, against a
realized $\max_t\|\q-\m^\star\|=7.27$ along the reference trajectories: the
guarantee is finite and explicit but, as expected of a bound that holds for all
inputs in the box and for all time, roughly seventeen times larger than the
excursion actually taken.

\paragraph{Why the plain certificate fails.}
That $\alpha<0$ does not contradict the continuum energy identity
\eqref{eq:burgersenergy}, nor is it an artefact of where the coordinates sit. The
basis is centered, so $\q=\mathbf 0$ corresponds to $\qq=\bar\qq$, the reference
state the ROM expands about, and the snapshot trajectory does orbit around
$\q=\mathbf 0$: over the training window the time average of $\q$ has norm $0.14$
against a root-mean-square $\|\q\|$ of $2.82$, the residual being the sampling gap
between the $m$ random observation times and the fine grid. The intrusive reduced model about the same $\bar\qq$ is dissipative,
$\alpha=+1.446$. The inferred $\Ahat$ is simply not an estimate of that operator:
the two differ by $99.8\%$ in norm, because the dissipation of $\mathbf{A}_F$ sits
in stiff modes the data never excite --- the intrusive $\sym\Ahat$ has eigenvalues
from $-4.3\times10^{3}$ to $-1.45$, the inferred one from $-10.9$ to $+0.77$ ---
and along the observed trajectory $\Ahat\q$ and $\Hhat(\q\otimes\q)$ are
identifiable only in combination. What the data determine is the behavior near
the states they visit, not the linear part in isolation.

The shift recovers exactly that combination. By \eqref{eq:shifted} the change of
variables $\s=\q-\m$ leaves $\Hhat$ untouched and replaces $\Ahat$ by
$\Ahat+\Lm_\m$, so $\alpha^\star$ certifies the operator the data actually
constrain. Write $\alpha_\m=-\lmax(\sym(\Ahat+\Lm_\m))$ for the decay rate certified at a
center $\m$. Since the basis is centered, $\q=\mathbf 0$ corresponds to $\qq=\bar\qq$, so the
plain certificate is already the one taken at the center the data suggest. At that
center the intrusive model is
dissipative, $\alpha_{\mathbf 0}=+1.446$, and the inferred one is not,
$\alpha_{\mathbf 0}=-0.771$. The semidefinite program finds a center at which it
is, returning $\alpha^\star=+0.364$ at $\|\m^\star\|=3.90$.

No certificate of \cref{sec:cert} applies to the unconstrained posterior, since
its cubic term does not vanish and cannot be dropped from $\dot V$; reporting
$\alpha$ for it would report a quantity that certifies nothing. The empirical
counterpart is that $23\%$ of its draws produce a reduced model that diverges
before $t=2$, against none of the certified draws.

\paragraph{Accuracy.}
\Cref{tab:prob1} reports the relative state error of the posterior-mean
prediction over $200$ draws, on the training window $[0,1]$ and on the
prediction window $]1,2]$ from which no data were taken, together with the
projection error of the full-order solution as the attainable floor.

\begin{table}[htbp]
\centering
\caption{Problem~1, $r=8$: relative state error of the ensemble mean over $200$
posterior draws. ``diverged'' counts the draws, out of $200$, whose reduced
model failed to integrate to $t=2$; those draws are excluded from the
unconstrained ensemble mean, which therefore flatters it.}
\label{tab:prob1}
\begin{tabular}{lcccccc}
\toprule
& \multicolumn{2}{c}{constrained} & \multicolumn{2}{c}{unconstrained}
& \multicolumn{2}{c}{projection floor} \\
\cmidrule(lr){2-3}\cmidrule(lr){4-5}\cmidrule(lr){6-7}
$(a,b)$ & $[0,1]$ & $]1,2]$ & $[0,1]$ & $]1,2]$ & $[0,1]$ & $]1,2]$ \\
\midrule
$(0.6,0.6)$   & $0.0080$ & $0.0620$ & $0.0082$ & $0.1214$ & $0.0078$ & $0.0322$ \\
$(1.4,0.6)$   & $0.0103$ & $0.0834$ & $0.0118$ & $0.2442$ & $0.0099$ & $0.0360$ \\
$(0.6,1.4)$   & $0.0116$ & $0.0702$ & $0.0163$ & $0.1207$ & $0.0113$ & $0.0279$ \\
$(1.4,1.4)$   & $0.0104$ & $0.0964$ & $0.0112$ & $0.2442$ & $0.0099$ & $0.0323$ \\
$(1.0,1.0)$   & $0.0068$ & $0.0711$ & $0.0092$ & $0.1625$ & $0.0065$ & $0.0309$ \\
\midrule
$(a^\star,b^\star)=(1.15,0.85)$ & $0.0071$ & $0.0745$ & $0.0092$ & $0.1938$ & $0.0069$ & $0.0325$ \\
\bottomrule
\end{tabular}

\vspace{2mm}
\begin{tabular}{lcccccc}
\toprule
$(a,b)$ & $(0.6,0.6)$ & $(1.4,0.6)$ & $(0.6,1.4)$ & $(1.4,1.4)$ & $(1.0,1.0)$ & $(a^\star,b^\star)$ \\
\midrule
diverged (unconstrained) & $2$ & $60$ & $54$ & $93$ & $35$ & $32$ \\
\bottomrule
\end{tabular}
\end{table}

\begin{figure}[htbp]
\centering
\includegraphics[width=\textwidth]{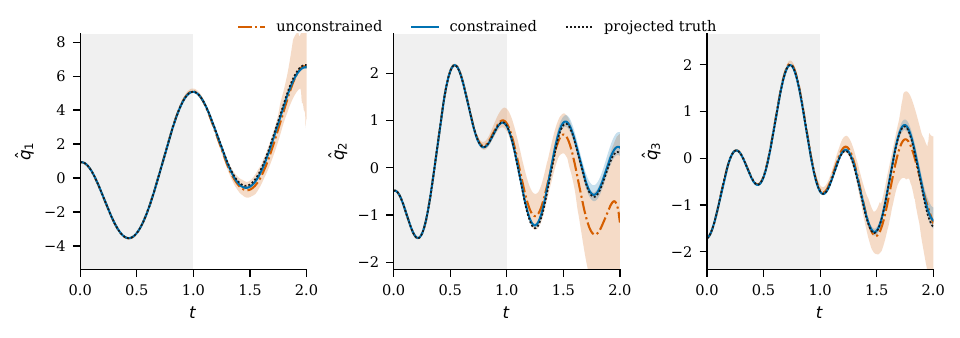}
\caption{Problem~1 at the held-out parameter value
$(a^\star,b^\star)=(1.15,0.85)$: the three leading reduced coordinates, posterior mean
and $95\%$ band over $200$ draws, against the projected
full-order solution. The shaded strip is the training window; no data were taken
from $t>1$. The axes are scaled to the truth and the constrained band, and the
unconstrained band leaves them as $t$ approaches $2$, where some of its surviving
draws spike; a further $32$ of its $200$ draws diverged outright and are absent
from the ensemble.}
\label{fig:prob1pred}
\end{figure}

On the training window both posteriors sit at the projection floor, as they
should: the data constrain the dynamics there and the constraint costs nothing.
The two separate on the prediction window, where the constrained posterior stays
within a factor of three of the floor while the unconstrained one is a further factor
of $1.7$ to $2.9$ worse --- and that comparison is generous to the baseline, since its
diverged draws are simply absent from its ensemble mean. The held-out parameter
value behaves like the training values for both, so the gap is not an
extrapolation-in-$(a,b)$ effect but an extrapolation-in-time one, which is what
\cref{fig:prob1pred} shows: the two posteriors are indistinguishable while data
are available and separate once the model has to carry itself.

\subsection{Problem~2: Korteweg--de Vries}
\label{sec:prob2}

The equation and its parameters are those of~\cite{Yildiz2025,EPrepr2025}. On the
periodic domain $x\in[0,P[$ with $P=2$, and with no forcing,
\begin{equation}
  \frac{\partial w}{\partial t}+\eta\,w\frac{\partial w}{\partial x}
  +\gamma^2\frac{\partial^3w}{\partial x^3}=0,
  \qquad
  w(x,0;a,b)=a\operatorname{sech}\bigl(b(x-P/2)\bigr),
  \label{eq:kdv}
\end{equation}
with $\gamma=0.022$ and $\eta=1$. The amplitude and width of the initial pulse are
the parameters: $(a,b)$ ranges over $[0.3,0.5]\times[0.8,1.2]$, with the four
corners and the center used for training and $(a^\star,b^\star)=(0.45,0.9)$ held out.
Observations are taken on $t\in[0,15]$, predictions carried to $t=30$, and
$r=10$, as in the reference. Only the equation is shared with those references: they infer from a
dense noise-free trajectory, whereas the data here are the $m=120$ noisy snapshots
per trajectory of the observation model above, so the results are not comparable
and no comparison is attempted.

\paragraph{Energy preservation and full-order discretization.} 
Three things distinguish this problem from the first, and they are the reason for
including it. There is no input, so $\Bhat$ is absent, the feature vector has
length $d=1+r+r(r+1)/2$, and the certified radius of \cref{thm:trap} reduces to
$R_\m=\|\mathbf{g}(\m)\|/\alpha_\m$.

There is also no dissipation. The convective contribution to
$E(t)=\tfrac12\int_0^Pw^2\,\mathrm dx$ vanishes by the argument of
\eqref{eq:burgersenergy}, which is unchanged here, and the dispersive one vanishes
too: integrating by parts twice on the circle,
\begin{equation}
  \gamma^2\int_0^Pw\,w_{xxx}\,\mathrm dx
  =-\gamma^2\int_0^Pw_xw_{xx}\,\mathrm dx
  =-\frac{\gamma^2}{2}\Bigl[w_x^2\Bigr]_0^P=0 .
  \label{eq:kdvenergy}
\end{equation}
So $\mathrm dE/\mathrm dt=0$: \eqref{eq:kdv} conserves energy exactly, where
\eqref{eq:burgers1d} merely dissipated it.

The convective term is discretized as in \eqref{eq:skewsym}, with the same
$\lambda=\tfrac23$ of Appendix~\ref{app:lambda}; the third derivative is discretized by the centered difference
\begin{equation}
  \bigl[\mathbf{A}_F\qq\bigr]_i
  =-\gamma^2\,\frac{w_{i+2}-2w_{i+1}+2w_{i-1}-w_{i-2}}{2\Delta x^3},
  \label{eq:kdvdisc}
\end{equation}
on a grid of $N=500$ points, $\Delta x=0.004$, advanced by the implicit midpoint
rule at $\Delta t=5\times10^{-3}$: the rule conserves every quadratic first
integral exactly, so the reference solution cannot leak the energy the reduced
models are judged on. What remains is rounding, which accumulates over the
$K=6\times10^{3}$ steps and grows like $\sqrt{K}\,\epsilon_{\mathrm{mach}}$; measured,
$|\Delta E/E|<10^{-13}$ over $[0,30]$ for every trajectory. On a uniform periodic grid the stencil is skew-symmetric, so $\sym\mathbf{A}_F=\mathbf 0$, and skewness survives
projection: an uncentered basis would give $\sym\Ahat=\mathbf 0$ and $\alpha=0$
exactly. Centering on $\bar\qq$ adds the mean-flow term
$2\Hm_F(\bar\qq\otimes\Id_N)$ to $\Ahat$, which is not skew. For the intrusive
operators at $r=10$ this gives $\alpha=-\lambda_{\max}(\sym\Ahat)=-8.2\times10^{-3}$;
minimizing $\lambda_{\max}(\sym(\Ahat+\Lm_\m))$ over $\m$ as in \cref{thm:trap}
improves it only to $\alpha_{\m^\star}=-1.6\times10^{-3}$, so no radius is
returned for the reference model.

Little should be read into that, in either direction. A trap is a set the flow is
drawn into and \eqref{eq:kdv} has none, being bounded by its initial energy
rather than by an attractor, so a radius obtained here would be a property of the
inferred model and not of the equation. Nor is certifying a coin toss that a
draw wins half the time: it asks all $r$ eigenvalues of $\sym\Ahat$ to be
non-positive together, and the reference model already has five of its ten
positive, spread as widely as the five negative ones. We report $\hat p_\delta$
and the certified radii for completeness and expect them to be uninformative.
What problem~2 is for is accuracy, in the format of \cref{tab:prob1}, and the
comparison is cleaner here than in problem~1: under the constraint a draw can gain
energy only through $\Ahat$, since $\q\T\Hhat(\q\otimes\q)$ vanishes identically,
whereas the unconstrained posterior has the second route as well, an $\Hhat$ that
feeds energy in. In problem~1 viscosity absorbs whatever $\Hhat$ produces and
hides the difference; here nothing does, so the drift of $E$ along the predicted
trajectories measures it directly.

\subsubsection{Results for problem~2}
\label{sec:results2}

\paragraph{Three posteriors.}
Problem~2 is where the linear part can be constrained as well, so three
posteriors are compared on the same data, the same basis and the same seeds: the
GP-BayesOpInf baseline, the energy-preserving posterior of \cref{sec:bayes}, and
the posterior supported on $\SEP\cap\SSK$ of \cref{sec:conspost}. They differ
only in $\C$, which has $n_c=220$ rows, then $n_c+n_a=275$. The prior scales are
selected separately for each by \cref{prop:evidence}. In the constrained
posteriors the structure is exact draw by draw: $\|\C\bth\|$ is at most
$2\times10^{-15}$ over $200$ draws, the relative energy residual of $\Hhat$ is
$4\times10^{-17}$ against $9\times10^{-2}$ for the baseline, and
$\|\sym\Ahat\|_F$ is $5\times10^{-16}$ under $\SSK$ against $0.49$ without it.

\paragraph{Certificates.}
As anticipated, nothing. For the energy-preserving posterior, over $3000$ draws
not one is certified by either test: $\hat p_0=0$, with Clopper--Pearson interval
$[0,\,1.2\times10^{-3}]$. The plain rate has median $-0.38$ and never exceeds
$-0.35$; the shifted rate improves this to a median of $-0.14$ but never reaches
zero either.

The reason is visible in the spectrum rather than in $\hat p_0$. Four of the
$r=10$ eigenvalues of $\sym\Ahat$ are positive in $99.8\,\%$ of the draws and five
in the rest, a mean of $4.0$: about half, which is what one expects of an operator
that is nearly skew-symmetric. A certificate needs all ten to be non-positive at
the same time, and no draw does that. The half-positive spectrum and the zero
probability are the same fact.

For the third posterior the certificates are not reported, because
\cref{rem:certvacuous} makes them empty before any data are seen: with
$\sym\Ahat=\mathbf{0}$ the rate $\alpha$ is zero by construction, which the
draws confirm to solver accuracy. What replaces them is \cref{thm:skew}. Here
$\|\chat\|=0.062$, so every draw must satisfy
$\|\q(t)\|\le\|\q_0\|+0.062\,t$, and all $60$ do at all six parameter values ---
a guarantee that needed no sampling to establish and admits no exceptions.

\paragraph{Accuracy.}
\Cref{tab:prob2} reports the relative state error of the posterior-mean prediction
over $60$ draws, split into the training window $[0,15]$ and the prediction
window $]15,30]$, with the projection error as the attainable floor.

\begin{table}[htbp]
\centering
\caption{Problem~2, $r=10$: relative state error of the ensemble mean over $60$
posterior draws. Every draw of both constrained posteriors integrates to $t=30$ at
every parameter value. The baseline does not: how many of its draws survive is
listed beneath the table, its error is computed over those only, and a dash marks a
parameter value at which it has no trajectory left to report.}
\label{tab:prob2}
\begin{tabular}{lcccccccc}
\toprule
& \multicolumn{2}{c}{$\Hhat\in\SEP$}
& \multicolumn{2}{c}{$\Hhat\in\SEP$, $\Ahat\in\SSK$}
& \multicolumn{2}{c}{unconstrained}
& \multicolumn{2}{c}{projection floor} \\
\cmidrule(lr){2-3}\cmidrule(lr){4-5}\cmidrule(lr){6-7}\cmidrule(lr){8-9}
$(a,b)$ & $[0,15]$ & $]15,30]$ & $[0,15]$ & $]15,30]$ & $[0,15]$ & $]15,30]$
& $[0,15]$ & $]15,30]$ \\
\midrule
$(0.30,0.8)$  & $0.0123$ & $0.0349$ & $0.0111$ & $0.0285$ & $0.0112$ & $0.0388$ & $0.0065$ & $0.0071$ \\
$(0.50,0.8)$  & $0.0233$ & $0.0752$ & $0.0220$ & $0.0592$ & ---      & ---      & $0.0090$ & $0.0107$ \\
$(0.30,1.2)$  & $0.0231$ & $0.1288$ & $0.0226$ & $0.0669$ & ---      & ---      & $0.0174$ & $0.0204$ \\
$(0.50,1.2)$  & $0.0611$ & $0.2418$ & $0.0606$ & $0.1248$ & $0.3888$ & $0.4529$ & $0.0318$ & $0.0338$ \\
$(0.40,1.0)$  & $0.0238$ & $0.1103$ & $0.0235$ & $0.0450$ & $0.1091$ & $0.3899$ & $0.0148$ & $0.0158$ \\
\midrule
$(0.45,0.9)^\star$ & $0.0220$ & $0.1004$ & $0.0214$ & $0.0489$ & $0.1847$ & $0.4593$ & $0.0114$ & $0.0143$ \\
\midrule
mean & $0.0276$ & $0.1152$ & $0.0269$ & $0.0622$ & & & & \\
\bottomrule
\end{tabular}

\medskip
\footnotesize Surviving baseline draws, of $60$: $60$, $0$, $0$, $13$, $15$, $14$.
\end{table}

The first thing the table settles is whether the model runs at all. Every draw of
both constrained posteriors reaches $t=30$ at every parameter value. The baseline
loses all $60$ at two of them and all but about a quarter at three more, and the
trajectories that do survive are the worse ones: on the prediction window its
error is up to five times the constrained one wherever it can be measured. In problem~1 the same comparison
cost the baseline $23\,\%$ of its draws and viscosity absorbed the rest; here,
with no dissipation to absorb anything, an $\Hhat$ that is not energy preserving
injects energy that nothing removes, and the state grows without bound.

The second thing it settles is what the linear constraint adds. On the observation
window the two constrained posteriors are indistinguishable, as they should be:
both fit the same derivative data, and $\SSK$ removes directions the data barely
inform. On the prediction window the skew-constrained posterior is better at every
parameter value, by a factor of $1.2$ to $2.5$, and its mean error is $6.2\,\%$
against $11.5\,\%$. The uncertainty tightens with it: averaged over $]15,30]$ the
width of the $95\,\%$ band on the first six coordinates of the held-out trajectory
falls from $0.006$--$0.121$ to $0.002$--$0.048$.

Both are to be read against the floor. The projection error at $r=10$ is
$0.7$--$3.4\,\%$, so the constrained models are within a factor of three of the basis
on the observed window and roughly a factor of three to seven beyond it; the remaining
error is the inference, not the basis.

\begin{figure}[htbp]
\centering
\includegraphics[width=\textwidth]{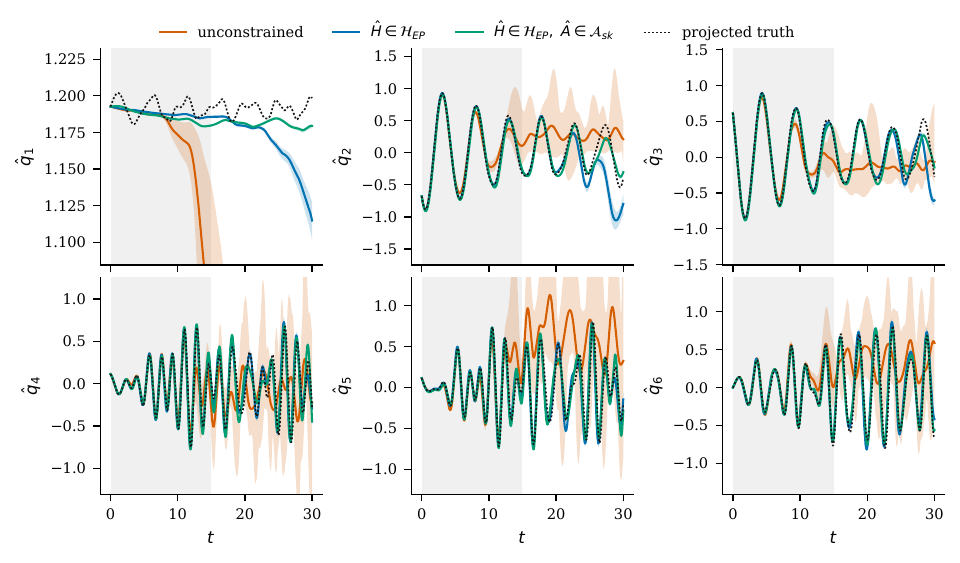}
\caption{Problem~2 at the held-out $(a^\star,b^\star)=(0.45,0.9)$: the first six
reduced coordinates of the three posteriors, each as its ensemble mean over $60$
draws with a $95\,\%$ band, against the projected reference solution (black).
The shaded strip is the training window $[0,15]$; everything to its right is
prediction. The vertical range of each panel is set from the reference solution
and the two constrained posteriors, so the baseline (orange) leaves the frame
where it diverges. Constraining $\Hhat$ alone (blue) keeps every draw bounded but
drifts in $\hat q_1$ and widens; constraining $\Ahat$ as well (green) tracks the
reference through the prediction window with a band about twice as narrow.}
\label{fig:prob2}
\end{figure}

\Cref{fig:prob2} shows the same comparison on the held-out trajectory. The three
posteriors are indistinguishable while the data last, and separate immediately
afterwards: the baseline leaves the scale of the problem, the energy-preserving
posterior keeps the oscillation but loses amplitude and drifts in the leading
coordinate, and the doubly constrained one stays on the reference with a band
that barely widens.

\paragraph{Energy.}
The reduced energy $V=\tfrac12\|\q\|^2$ is computed along each predicted
trajectory separately and summarized over the $60$ draws at each parameter value.
Three numbers are reported, in each case for the energy-preserving posterior
first and for the doubly constrained one second. The median draw changes $V$ over
$[0,30]$ by between $7\,\%$ and a factor of $2.6$, depending on the parameter
value, against between $1\,\%$ and $7\,\%$. The worst single draw changes it by a
factor of $5.0$, against $7\,\%$. The largest $\|\q\|$ reached along a trajectory
is $2.5$ times the initial value, against $1.02$ times. For comparison, the reference solution is
constant to round-off. The baseline cannot be summarized in the same way, because
at two of the six parameter values none of its draws survive to $t=30$; at the four
where some do, its energy changes by between $0.6\,\%$ and $19\,\%$. This is \cref{thm:skew} in numbers: once $\q\T\Hhat(\q\otimes\q)$
and $\q\T\Ahat\q$ both vanish identically, the only term left in the energy
balance is $\q\T\chat$, and here $\|\chat\|=0.062$.

Structure can be imposed on the linear part in other ways. A physics-informed
regularizer penalizes departures from the desired structure without ruling them
out~\cite{Sawant2023}; a dissipative parametrization rules them out exactly, by
writing $\Ahat$ in a form that admits only dissipative
operators~\cite{Goyal2025,Yildiz2025}. The constraint used here is exact, like the
second, but is linear in the operator coefficients, like the first. That
combination is what matters for a Bayesian method: the posterior stays Gaussian
and the closed forms of \cref{sec:bayes} apply unchanged, which a nonlinear
parametrization of $\Ahat$ would not allow. What it costs is the certificate,
which \cref{rem:certvacuous} makes empty. For \eqref{eq:kdv} that is the right
trade, since the equation has no attracting set to certify in the first place.

\subsection{Problem~3: two-dimensional Burgers}
\label{sec:prob3}

The third problem is two-dimensional, and it is viscous, so the certificates of
\cref{sec:cert} have something to certify again after problem~2. The equation and
its coefficients are those of~\cite{EPrepr2025}. On the periodic unit square
$(x,y)\in[0,1[^2$,
\begin{equation}
  \frac{\partial w}{\partial t}
  =\kappa\,w\Bigl(\frac{\partial w}{\partial x}+\frac{\partial w}{\partial y}\Bigr)
  +\nu\Bigl(\frac{\partial^2w}{\partial x^2}+\frac{\partial^2w}{\partial y^2}\Bigr),
  \label{eq:burgers2d}
\end{equation}
with $\kappa=0.2$ and $\nu=2\times10^{-3}$, the convective coefficient being
written $\kappa$ so that $c$ remains free for the constant term of the ROM. The
state is a scalar field, as in problems~1 and~2; what is new is that it is
transported in two directions at once. The initial condition is
\begin{equation}
  w(x,y,0;a,b)=a\bigl(\cos2\pi x\,\cos2\pi y
    +b\cos4\pi x\,\cos4\pi y\bigr),
  \label{eq:burgers2dic}
\end{equation}
whose amplitude $a$ and second-harmonic weight $b$ are the parameters; the
reference takes $a=1$ and $b=0$. As in problem~2, the parameters enter through
the initial state. The two harmonics interact as the profile steepens, so $b$
changes the dynamics rather than merely rescaling it.

\paragraph{Energy preservation and full-order discretization.}
With $E(t)=\tfrac12\int_0^1\!\!\int_0^1w^2\,\mathrm dx\,\mathrm dy$,
\begin{equation}
  \frac{\mathrm dE}{\mathrm dt}
  =\kappa\int\!\!\int w^2\Bigl(\frac{\partial w}{\partial x}
     +\frac{\partial w}{\partial y}\Bigr)\mathrm dx\,\mathrm dy
   +\nu\int\!\!\int w\,\Delta w\,\mathrm dx\,\mathrm dy
  =-\nu\int\!\!\int\|\nabla w\|^2\,\mathrm dx\,\mathrm dy .
  \label{eq:burgers2denergy}
\end{equation}
The convective contribution vanishes one direction at a time: integrating in $x$
first, $\int_0^1w^2w_x\,\mathrm dx=[w^3/3]_0^1=0$ by periodicity, and the same
argument in $y$ disposes of the second term. It is the identity of
\eqref{eq:burgersenergy} applied twice, and it needs nothing of $w$ beyond
periodicity. Viscosity dissipates, and with no forcing these are the only two
contributions, so $E$ decreases monotonically at a rate set by $\nu$.

On a uniform $N\times N$ grid with $\Delta x=\Delta y=1/N$ and periodic indices,
the Laplacian is the five-point stencil and each convective derivative is the
$\lambda=\tfrac23$ stencil \eqref{eq:skewsym} of problem~1
(Appendix~\ref{app:lambda}), applied in its own direction. The argument of \cref{sec:prob1} carries over term by term, so
$\qq\T\Hm_F(\qq\otimes\qq)=0$ for every $\qq$; we again verify it on
random states, which matters more here, since on a symmetric field such as
\eqref{eq:burgers2dic} the violation of a non-conservative stencil cancels and
goes unnoticed. The viscous part gives $\mathbf{A}_F=\nu\mathbf{L}$ with
$\mathbf{L}$ the five-point Laplacian, so $\sym(\mathbf{A}_F)\preceq\mathbf{0}$.

Two consequences shape the experiment: \cref{prop:galerkin} applies exactly, so
the hard constraint of \cref{sec:bayes} is the appropriate one; and
$\sym(\mathbf{A}_F)\neq\mathbf{0}$, so the constraint of \cref{sec:conservative}
does \emph{not} apply, since imposing it would delete the viscosity.

\paragraph{Setup.}
Nominal values are those of the reference: $\Delta x=\Delta y=0.02$, so $N=50$
and $N_{\mathrm{dof}}=2500$, and a final time $T=4$. The parameters $(a,b)$ range over
$[0.4,0.6]\times[0,0.4]$, with the four corners and the center used for training
and $(a^\star,b^\star)=(0.55,0.1)$ held out. Observations are taken on $t\in[0,2]$
and predictions carried to $t=4$, with $m=30$ snapshots per trajectory at
$\xi=3\,\%$ noise, and $r=6$. Time integration is again
by the implicit midpoint rule, here for accuracy over the prediction window
rather than for exact conservation, since the viscous term makes the reference
energy decay.

\subsubsection{Results for problem~3}
\label{sec:results3}

\paragraph{Exactness.}
The same picture as in problem~1: $n_c=56$ rows of full rank, $\dim\ker\C=112$,
$\|\C\bth_k\|\le2.5\times10^{-14}$ over $200$ draws and
$\|\C\bmu\|=2.3\times10^{-14}$ at the mean against $1.48$ without the constraint,
and a relative energy residual of $\Hhat$ of $1.0\times10^{-15}$ against $0.109$
for the baseline. The prior scales of \cref{prop:evidence} are
$(\gamma_c,\gamma_A,\gamma_H)=(0.749,2.098,7.312)$ with the constraint and
$(0.533,1.319,6.046)$ without it, with log evidences $4325.7$ and $5141.4$.

\paragraph{Certificates.}
In problem~2 the certificates were empty. Here they work, but only in their shifted
form. The plain test \eqref{eq:alphaK} fails on all $3000$ draws: $\alpha$ has
median $-0.684$ and never exceeds $-0.472$, because two of the six eigenvalues of
$\sym\Ahat$ are positive in every draw. The shifted test \eqref{eq:sdp} succeeds on
all of them,
\begin{equation*}
  \hat p_\delta = 1.0000,
  \qquad
  \text{Clopper--Pearson } 95\%:\ [0.9988,\ 1],
\end{equation*}
with $\alpha^\star$ between $0.189$ and $0.261$, median $0.215$, and a certified
radius of median $R=15.4$, ranging from $6.3$ to $56.4$. At the posterior mean
$\alpha=-0.677$ and $\alpha^\star=+0.216$ at $\|\m^\star\|=14.1$, so $R=20.1$.

\paragraph{Accuracy.}
\Cref{tab:prob3} reports the relative state error of the posterior-mean prediction
over $200$ draws, on the training window $[0,2]$ and the prediction window
$]2,4]$, with the projection error as the attainable floor.

\begin{table}[htbp]
\centering
\caption{Problem~3, $r=6$: relative state error of the ensemble mean over $200$
posterior draws. Every constrained draw integrates to $t=4$ at every parameter
value; the baseline loses $377$ of its $1200$, its error is computed over the
survivors only, and how many survive is listed beneath the table.}
\label{tab:prob3}
\begin{tabular}{lcccccc}
\toprule
& \multicolumn{2}{c}{constrained} & \multicolumn{2}{c}{unconstrained}
& \multicolumn{2}{c}{projection floor} \\
\cmidrule(lr){2-3}\cmidrule(lr){4-5}\cmidrule(lr){6-7}
$(a,b)$ & $[0,2]$ & $]2,4]$ & $[0,2]$ & $]2,4]$ & $[0,2]$ & $]2,4]$ \\
\midrule
$(0.40,0.0)$ & $0.0144$ & $0.0550$ & $0.0153$ & $0.0962$ & $0.0144$ & $0.0452$ \\
$(0.60,0.0)$ & $0.0179$ & $0.1656$ & $0.0821$ & $0.5803$ & $0.0177$ & $0.1517$ \\
$(0.40,0.4)$ & $0.0137$ & $0.0596$ & $0.0142$ & $0.1167$ & $0.0135$ & $0.0528$ \\
$(0.60,0.4)$ & $0.0229$ & $0.1796$ & $0.0232$ & $0.2239$ & $0.0228$ & $0.1748$ \\
$(0.50,0.2)$ & $0.0122$ & $0.0946$ & $0.0127$ & $0.1691$ & $0.0121$ & $0.0865$ \\
\midrule
$(0.55,0.1)^\star$ & $0.0138$ & $0.1243$ & $0.0318$ & $10.63$ & $0.0134$ & $0.1146$ \\
\midrule
mean & $0.0158$ & $0.1131$ & $0.0299$ & $1.9689$ & $0.0156$ & $0.1043$ \\
\bottomrule
\end{tabular}

\medskip
\footnotesize Surviving baseline draws, of $200$: $200$, $76$, $200$, $133$, $178$,
$36$.
\end{table}

\begin{figure}[htbp]
\centering
\includegraphics[width=\textwidth]{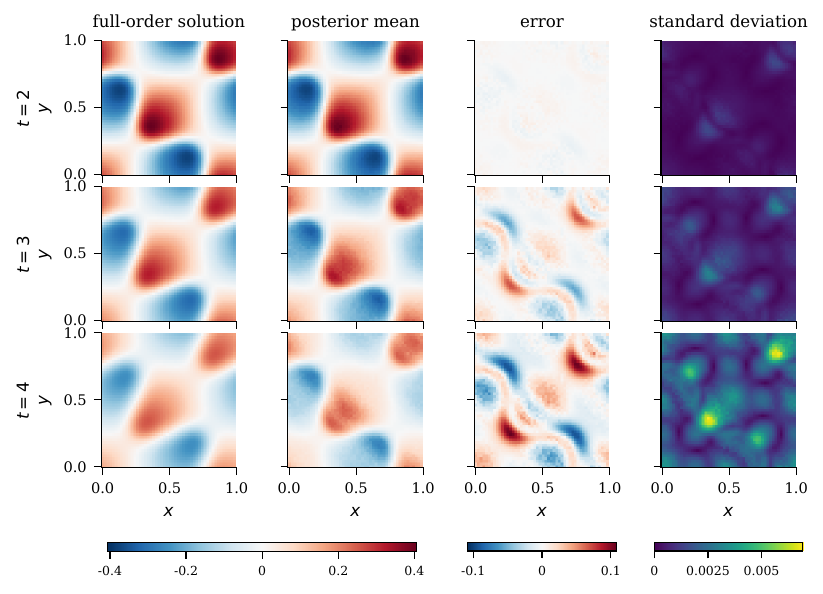}
\caption{Problem~3 at the held-out $(a^\star,b^\star)=(0.55,0.1)$: the field at the
end of the training window and at the two later times the model reaches on its
own. Columns: the full-order solution, the posterior mean over $200$ draws lifted
back to the $2500$ nodes, the error, and the pointwise standard deviation. The
first two columns share a color scale. The relative field error is $0.013$ at
$t=2$, $0.116$ at $t=3$ and $0.249$ at $t=4$, where the projection floor alone is
$0.19$.}
\label{fig:prob3fields}
\end{figure}

The constrained posterior is better than the baseline at every parameter value and
on both windows. On the training window it sits on the projection floor to the
third decimal --- $0.0144$ against $0.0144$, $0.0229$ against $0.0228$ --- so the
inference costs nothing the basis has not already spent. On the prediction window
it stays within $10\,\%$ of the floor at four of the six parameter values and
within a quarter of it at the other two, while the baseline is a factor of $1.3$ to
$3.5$ worse wherever its draws survive; at the held-out value only $36$ of its
$200$ draws survive at all, and those average an error above $10$.
\Cref{fig:prob3fields} shows what that means as a field: at $t=2$ the error is
invisible on the scale of the solution, and by $t=4$ it is concentrated where the
profile has steepened.

The bands are narrow: the median width of the $95\,\%$ interval over the reduced
coordinates is $0.006$ -- $0.020$ on $[0,2]$ and $0.025$ -- $0.116$ on $]2,4]$. The
posterior describes its uncertainty about the operator, and at $r=6$ that
uncertainty is small next to the closure error of the basis, which it does not
model; the projection floor in \cref{tab:prob3} is where most of the remaining
error sits.

\paragraph{Energy.}
The full-order energy decreases monotonically, and so does the constrained
posterior: all $1200$ of its lifted trajectories lose energy at every step, at
every parameter value, their largest one-step change being $-1.6$ to
$-2.0\times10^{-3}$ relative, against $-1.4\times10^{-3}$ for the reference
solutions. The baseline gains energy instead. Between $31$ and $86\,\%$ of its
surviving draws gain some at one time or another, depending on the parameter
value: $102$ of $200$ at $(0.40,0)$, and $31$ of $36$ at the held-out value. The
largest gain is $2.7\times10^{6}$ times the initial energy. The reason is the one
the constraint was imposed for. Under it $\q\T\Hhat(\q\otimes\q)=0$, so the
quadratic term moves no energy and only the linear part can change it; the
certificate above says that part decays, about the center $\m^\star$. Without the
constraint the quadratic term moves energy too, and nothing bounds how much it
adds.

\section{Conclusions and outlook}
\label{sec:conclusion}

The property that makes the reduced energy a global Lyapunov function for a
quadratic ROM, i.e.,  energy preservation of the quadratic operator, is a set of linear
equality constraints on the operator coefficients. A linear constraint is
exactly what a Gaussian posterior can absorb, so the structure can be imposed
without approximation and without violating closed form. The constrained
posterior is one Kalman-type update of the unconstrained one, and exact samples
follow by a linear correction, while the constraint survives later linear-Gaussian
updates. The same object has three equivalent descriptions: a degenerate
prior on the energy-preserving subspace, conditioning on a noise-free
pseudo-observation, and constrained maximum a posteriori estimation with its
reduced Hessian. A soft variant for imposing the constraints is an anisotropic penalty whose scale is selected
by closed-form evidence maximization and controls a certified local stability
region.

What the structure buys for the dynamics is that global boundedness and
input-to-state stability reduce to one linear matrix condition on $\sym\Ahat$, potentially 
with a convex shift of the energy covering linear parts that are not
dissipative. This makes stability assessment a number: the posterior probability of
certified global boundedness, estimated with exact confidence bounds and used to
sample the certified posterior exactly, which replaces the empirical finite-horizon
simulation test in OpInf practice. When the full-order system conserves
energy instead of dissipating it, constraining the linear part to be skew is an additional
constraint of the same kind. In this case $\sym\Ahat$ vanishes, and the certificate is empty
by construction, with a deterministic linear growth bound available for every
draw without sampling.

The experiments bear this out. Every constrained draw integrates stably to the
final time at every parameter value, in all three problems, where the
unconstrained baseline loses up to all of its draws, and beyond the observation
window the constrained posterior is the more accurate of the two everywhere.
Where the equation admits a decay rate the shifted certificate is informative,
holding for $95\,\%$ of the draws in problem~1 and for all of them in
problem~3; where it does not, the skew constraint takes its place and nearly
halves the prediction error on Korteweg--de Vries.

A few limitations are worth stating plainly. a) The certificate is sufficient, yet
not necessary, and what it certifies may not be the physical energy.
In problem~1, for example, the continuum energy already decays by \eqref{eq:burgersenergy},
yet the inferred $\sym\Ahat$ is not negative definite, and the test succeeds
only after the energy is recentered at an $\m^\star$ far from the snapshot mean,
for which we have no physical reading. A linear part that is stable but far from
normal --- eigenvalues in the left half-plane, yet $\sym\Ahat$ indefinite, so
that $\|\q\|$ grows before it decays --- may need a weighted
$V=\tfrac12\q\T\boldsymbol{\Xi}\q$ instead. b) The constraint belongs
to systems that preserve energy; on a system that does not, it is the wrong
model. c) Imposing the stability constraints also couples the row-wise regressions, so the cost grows faster in
$r$. d) The posterior describes the operators, not the basis: at $t=4$ in
problem~3 the projection error is $0.19$ and the total error $0.25$.

Future work can learn the energy metric itself, replacing
$V=\tfrac12\|\q\|^2$ by $V=\tfrac12\q\T\boldsymbol{\Xi}\q$ with $\boldsymbol{\Xi}\succ0$ symmetric, so
that energy is measured in a weighted inner product rather than the Euclidean
one. Energy preservation then reads $\q\T\boldsymbol{\Xi}\Hhat(\q\otimes\q)=0$, which is
bilinear in $(\boldsymbol{\Xi},\Hhat)$ rather than linear in $\Hhat$. The linear part may
likewise be constrained to a structured form. It is also meaningful to move to online settings,
where \cref{prop:closure} keeps energy preservation exact under sequential
updating.

\section*{CRediT authorship contribution statement}

\noindent\textbf{Jo\~ao Victor Galv\~ao da Mata:} Conceptualization, Investigation, Methodology, Software, Visualization, Validation,
Writing -- original draft, Writing -- review and editing.
\textbf{Mengwu Guo:} Conceptualization, Funding acquisition, Investigation,  Methodology, Supervision, Writing -- review and editing.

\section*{Declaration of competing interest}

The authors declare that they have no known competing financial interests or
personal relationships that could have appeared to influence the work reported
in this paper.

\section*{Acknowledgments}

The authors acknowledge financial support from the Swedish Research Council (VR) under No. 2025-04911 and from The Crafoord Foundation under No. 20250657.

\section*{Data availability}

The code and data in this work are publicly available at
\url{https://github.com/Jvictormata/EP_Gp_Bayes_Opinf}.

\section*{Declaration of generative AI and AI-assisted technologies in the manuscript preparation process}

During the preparation of this work, the authors used ChatGPT, Claude, and
Grammarly for grammar check and language polishing, especially to improve
textual clarity. After using these tools, the authors reviewed and edited the
content as needed and take full responsibility for the content of this article.

\appendix
\section{Proofs and derivations}
\label{app:proofs}

\subsection{Proof of \texorpdfstring{\Cref{prop:index}}{Proposition 3.2}}
\label{app:index}

$p$ is a homogeneous cubic polynomial, so it vanishes identically if and only if
all its monomial coefficients vanish. The coefficient of $\hat q_a\hat q_b\hat q_c$ is
the sum of $H_{ijk}$ over the \emph{distinct} rearrangements $(i,j,k)$ of
$(a,b,c)$. Summing instead over all six orderings and using \eqref{eq:symH}
gives
\[
  H_{abc}+H_{acb}+H_{bac}+H_{bca}+H_{cab}+H_{cba}=2\,(H_{abc}+H_{bac}+H_{cab}).
\]
If $a,b,c$ are distinct, the true coefficient is this sum. If exactly two
indices coincide, it is half the sum. If all coincide, it is one sixth of it.
In every case the coefficient is a positive multiple of
$H_{abc}+H_{bac}+H_{cab}$, which proves \eqref{eq:EPindex}. \qed

\subsection{Proof of \texorpdfstring{\Cref{lem:identity}}{Lemma 4.1}}
\label{app:identity}

Let $\mathbf{S}=\bSig^{1/2}$. The left-hand side of \eqref{eq:identity} is
$\mathbf{S}(\Id-\Pm_1)\mathbf{S}$, where $\Pm_1$ is the orthogonal projector onto
$\range(\mathbf{S}\C\T)$. Thus $\Id-\Pm_1$ is the orthogonal projector onto
$\range(\mathbf{S}\C\T)^\perp=\ker(\C\mathbf{S})=\mathbf{S}^{-1}\ker\C=\range(\mathbf{S}^{-1}\Tm)$.
The right-hand side equals $\mathbf{S}\Pm_2\mathbf{S}$ with
$\Pm_2=\mathbf{S}^{-1}\Tm(\Tm\T\mathbf{S}^{-2}\Tm)^{-1}\Tm\T\mathbf{S}^{-1}$, the orthogonal
projector onto $\range(\mathbf{S}^{-1}\Tm)$. Hence $\Pm_2=\Id-\Pm_1$. \qed

\subsection{Proof of \texorpdfstring{\Cref{thm:equiv}}{Theorem 4.2}}
\label{app:equiv}

(i) Apply \eqref{eq:kalman} and its limit with $\bSig_0$ in place of $\bSig$
and prior mean $\mathbf{0}$. The innovation vanishes, since both the prior mean
and the observed value of \eqref{eq:pseudo} are zero, so the conditioned mean is
$\mathbf{0}$. The conditioned covariance is
$\bSig_0-\bSig_0\C\T(\C\bSig_0\C\T)^{-1}\C\bSig_0=\Tm\bSig_{\varphi,0}\Tm\T$
by \eqref{eq:identity}, which is the first claim. For the commutation, fix $\epsilon>0$. Since $\bet$
is independent of the data, the two likelihoods factorize,
$p(\zt,\mathbf{0}\mid\bth)=p(\zt\mid\bth)\,p(\mathbf{0}\mid\bth)$, so the
posterior density is proportional to
$p(\zt\mid\bth)\,p(\mathbf{0}\mid\bth)\,p_0(\bth)$ whatever the order in which
the factors are absorbed; the two routes therefore agree for every
$\epsilon>0$. Both are Gaussian computations whose means and covariances are
continuous in $\epsilon$ at $0$, so they agree in the limit. Along the lower
route the data update must be written in covariance form, since the conditioned
prior is singular; it remains continuous because
$\Phim\bSig'\Phim\T+\W^{-1}\succeq\W^{-1}\succ0$ for every $\bSig'\succeq0$.
(ii) By the choice of $\bSig_{\varphi,0}$, the two terms in \eqref{eq:phipost}
carry the same sandwich and merge:
\[
  \bSig_\varphi^{-1}=\Tm\T\Phim\T\W\Phim\Tm+\Tm\T\bSig_0^{-1}\Tm
  =\Tm\T\bigl(\Phim\T\W\Phim+\bSig_0^{-1}\bigr)\Tm=\Tm\T\bSig^{-1}\Tm .
\]
Hence $\Tm\bSig_\varphi\Tm\T=\Tm(\Tm\T\bSig^{-1}\Tm)^{-1}\Tm\T=\bSig_c$ by
\eqref{eq:identity}, which is \eqref{eq:sigmacT}. Its mean is
$\Tm\bmu_\varphi=\Tm(\Tm\T\bSig^{-1}\Tm)^{-1}\Tm\T\bSig^{-1}\bmu$, using
$\Phim\T\W\zt=\bSig^{-1}\bmu$. This equals $\bmu_c$, because both are the
$\bSig^{-1}$-orthogonal projection of $\bmu$ onto $\ker\C$.
(iii) Completing the square, $J(\bth)=(\bth-\bmu)\T\bSig^{-1}(\bth-\bmu)+\text{const}$,
so $\nabla^2(\tfrac12 J)=\bSig^{-1}$ and $J$ is strictly convex; with linear
constraints the KKT conditions
$\bSig^{-1}(\bth-\bmu)+\C\T\bm\lambda=\mathbf{0}$, $\C\bth=\mathbf{0}$ are
necessary and sufficient, and give $\bm\lambda=(\C\bSig\C\T)^{-1}\C\bmu$ and
$\bth=\bmu_c$. The reduced Hessian is $\Tm\T\bSig^{-1}\Tm$; apply
\eqref{eq:identity}.
(iv) By direct computation from \eqref{eq:cond}, using
$\C\K_0=\Id_{n_c}$; see Appendix~\ref{app:kalman}. Then $\C\bth$ has mean $\mathbf{0}$
and covariance $\C\bSig_c\C\T=\mathbf{0}$, hence vanishes almost surely.
\qed

\subsection{Derivation of the pseudo-observation update}
\label{app:kalman}

We derive \eqref{eq:kalman} from \eqref{eq:pseudo}, verify that it is the
conditional law of $\bth$ given $\y=\mathbf{0}$, and justify the limit
\eqref{eq:cond}. Throughout $\bth\sim\Nor(\bmu,\bSig)$ with $\bSig\succ0$,
$\bet\sim\Nor(\mathbf{0},\epsilon^2\Id_{n_c})$ is independent of $\bth$, and
$\y=\C\bth+\bet$.

The pair $(\bth,\y)$ is jointly Gaussian, since
\[
  \begin{bmatrix}\bth\\ \y\end{bmatrix}
  =\begin{bmatrix}\Id_n&\mathbf{0}\\ \C&\Id_{n_c}\end{bmatrix}
   \begin{bmatrix}\bth\\ \bet\end{bmatrix}
\]
is an affine map of the Gaussian vector $(\bth,\bet)$. Its moments are
$\E[\y]=\C\bmu$,
\[
  \operatorname{Cov}(\bth,\y)
  =\E\bigl[(\bth-\bmu)\bigl(\C(\bth-\bmu)+\bet\bigr)\T\bigr]=\bSig\C\T,
  \qquad
  \operatorname{Cov}(\y)=\C\bSig\C\T+\epsilon^2\Id=:\mathbf{S}_\epsilon ,
\]
the cross term vanishing by independence of $\bth$ and $\bet$. Hence
\begin{equation}
  \begin{bmatrix}\bth\\ \y\end{bmatrix}\sim
  \Nor\left(\begin{bmatrix}\bmu\\ \C\bmu\end{bmatrix},
  \begin{bmatrix}\bSig&\bSig\C\T\\ \C\bSig&\mathbf{S}_\epsilon\end{bmatrix}\right).
  \label{eq:jointpseudo}
\end{equation}
The matrix $\mathbf{S}_\epsilon$, the \emph{innovation covariance}, is positive
definite: $\C$ has full row rank and $\bSig\succ0$, so
$\bm\lambda\T\C\bSig\C\T\bm\lambda>0$ for $\bm\lambda\neq\mathbf{0}$.

Since $\mathbf{S}_\epsilon$ is invertible, the standard conditioning formula
for a jointly Gaussian pair applied to \eqref{eq:jointpseudo} gives
\begin{equation}
  \bth\mid\y\ \sim\ \Nor\bigl(\bmu+\K_\epsilon(\y-\C\bmu),\
  \bSig-\K_\epsilon\C\bSig\bigr),
  \qquad \K_\epsilon:=\bSig\C\T\mathbf{S}_\epsilon^{-1} .
  \label{eq:condgen}
\end{equation}
Substituting the observed value $\y=\mathbf{0}$, the innovation is
$\mathbf{0}-\C\bmu=-\C\bmu$ and we obtain \eqref{eq:kalman}.

Equivalently, from Bayes' rule,
\[
  -\log p(\bth\mid\y=\mathbf{0})
  =\tfrac12(\bth-\bmu)\T\bSig^{-1}(\bth-\bmu)+\tfrac1{2\epsilon^2}\|\C\bth\|^2+\text{const},
\]
so that, completing the square,
$\bSig_\epsilon=(\bSig^{-1}+\epsilon^{-2}\C\T\C)^{-1}$ and
$\bmu_\epsilon=\bSig_\epsilon\bSig^{-1}\bmu$. This agrees with \eqref{eq:kalman}
by the Sherman--Morrison--Woodbury identity
$(\bSig^{-1}+\epsilon^{-2}\C\T\C)^{-1}=\bSig-\bSig\C\T(\C\bSig\C\T+\epsilon^2\Id)^{-1}\C\bSig$,
and, for the mean, by
$\bSig_\epsilon\bSig^{-1}\bmu=(\bSig-\K_\epsilon\C\bSig)\bSig^{-1}\bmu=\bmu-\K_\epsilon\C\bmu$.
This is the form used in \cref{prop:soft}: the pseudo-observation is the
penalty $\epsilon^{-2}\|\C\bth\|^2$. Note that the covariance form
\eqref{eq:kalman} admits the limit below, whereas the precision
$\bSig^{-1}+\epsilon^{-2}\C\T\C$ diverges.

The limit $\epsilon\to0$ exists because $\mathbf{S}_0=\C\bSig\C\T$ is
invertible, as shown above. Then $\mathbf{S}_\epsilon$ is invertible on a
neighborhood of $\epsilon=0$ and $\epsilon\mapsto\mathbf{S}_\epsilon^{-1}$ is
continuous there, matrix inversion being continuous on the open set of
invertible matrices. Hence $\K_\epsilon\to\K_0=\bSig\C\T(\C\bSig\C\T)^{-1}$ and
\eqref{eq:kalman} converges to \eqref{eq:cond}.

At $\epsilon=0$ we have $\C\K_0=(\C\bSig\C\T)(\C\bSig\C\T)^{-1}=\Id_{n_c}$,
from which \cref{thm:equiv}(iv) is immediate:
\[
  \C\bmu_c=\C\bmu-(\C\K_0)\C\bmu=\mathbf{0},\qquad
  \C\bSig_c=\C\bSig-(\C\K_0)\C\bSig=\mathbf{0}.
\]
The same identity shows that $\bm\Pi:=\Id-\K_0\C$ is idempotent,
$\bm\Pi^2=\Id-2\K_0\C+\K_0(\C\K_0)\C=\bm\Pi$, with $\C\bm\Pi=\mathbf{0}$ and
$\operatorname{rank}\bm\Pi=n-n_c$, so $\bm\Pi$ is the projection onto $\ker\C$
along $\range(\bSig\C\T)$, and
$\bmu_c=\bm\Pi\bmu$, $\bSig_c=\bm\Pi\bSig=\bm\Pi\bSig\bm\Pi\T$. The last
expression is the preferred one numerically and shows $\bSig_c\succeq0$
explicitly. Finally, for any $\mathbf{v}\in\R^n$,
\begin{equation}
  \mathbf{v}\T(\bSig-\bSig_\epsilon)\mathbf{v}
  =(\C\bSig\mathbf{v})\T\mathbf{S}_\epsilon^{-1}(\C\bSig\mathbf{v})\ \ge\ 0 ,
  \label{eq:varred}
\end{equation}
with equality if and only if $\C\bSig\mathbf{v}=\mathbf{0}$: the update reduces
the posterior variance in direction $\mathbf{v}$ exactly when $\mathbf{v}$ is
correlated with the constrained directions. This is the quantitative form of
the observation in \cref{rem:interp} that the EP update also sharpens, and
moves, the entries of $\Ahat$.

\subsection{Proof of \texorpdfstring{\Cref{prop:evidence}}{Proposition 4.10}}
\label{app:evidence}

With $n=\dim\bth$,
$p(\zt)=\int\Nor(\zt;\Phim\bth,\W^{-1})\,\Nor(\bth;\mathbf{0},\Pm_0^{-1})\,d\bth$.
Expanding the exponent,
\[
  (\zt-\Phim\bth)\T\W(\zt-\Phim\bth)+\bth\T\Pm_0\bth
  =(\bth-\bmu)\T\Pm(\bth-\bmu)-\bmu\T\Pm\bmu+\zt\T\W\zt .
\]
The normalizing constants contribute
$(2\pi)^{-(N+n)/2}|\W|^{1/2}|\Pm_0|^{1/2}$, and the Gaussian integral over
$\bth$ gives $(2\pi)^{n/2}|\Pm|^{-1/2}$. Taking logarithms yields
\eqref{eq:evidence}. \qed

\subsection{The energy-preserving choice \texorpdfstring{$\lambda=\tfrac23$}{lambda = 2/3}}
\label{app:lambda}

In the notation of \cref{sec:prob1}, since $\partial_x(w^2)=2ww_x$ the
continuum identity
\begin{equation}
  w\,w_x
  =\lambda\Bigl(\tfrac12\,\partial_x\bigl(w^2\bigr)\Bigr)
  +(1-\lambda)\,w\,w_x
  \qquad\text{for every }\lambda
  \label{eq:lambda}
\end{equation}
holds, which allows a split discretization. Applying centered differences to the
two terms on the right gives the one-parameter family
\begin{equation}
  \bigl[\Hm_F^{(\lambda)}(\qq\otimes\qq)\bigr]_i
  =-\lambda\,\frac{w_{i+1}^2-w_{i-1}^2}{4\Delta x}
   -(1-\lambda)\,w_i\,\frac{w_{i+1}-w_{i-1}}{2\Delta x},
  \label{eq:family}
\end{equation}
every member of which is second-order accurate and consistent with $-ww_x$:
\eqref{eq:lambda} is an identity, so it selects nothing in the continuum. The
discrete energy does.

Periodicity permits shifts of the summation index. Define
$T=\sum_iw_iw_{i+1}(w_{i+1}-w_i)$. Then
\begin{equation}
\begin{aligned}
  \sum_iw_i\bigl(w_{i+1}^2-w_{i-1}^2\bigr)
  &=\sum_iw_iw_{i+1}^2-\underbrace{\sum_iw_iw_{i-1}^2}_{\textstyle=\;\sum_iw_{i+1}w_i^2}
  \;=\;\sum_iw_iw_{i+1}\bigl(w_{i+1}-w_i\bigr)\;=:\;T,\\[4pt]
  \sum_iw_i^2\bigl(w_{i+1}-w_{i-1}\bigr)
  &=\sum_iw_i^2w_{i+1}-\underbrace{\sum_iw_i^2w_{i-1}}_{\textstyle=\;\sum_iw_{i+1}^2w_i}
  \;=\;-\sum_iw_iw_{i+1}\bigl(w_{i+1}-w_i\bigr)\;=\;-T .
\end{aligned}
\label{eq:biases}
\end{equation}
Substitution into \eqref{eq:family} gives the nonlinear contribution to the
discrete energy balance
\begin{equation}
  \qq\T\Hm_F^{(\lambda)}(\qq\otimes\qq)=-\frac{(3\lambda-2)\,T}{4\Delta x}.
  \label{eq:lamdefect}
\end{equation}
The quantity $T$ is not identically zero; for example, the mean-zero state
$(w_0,w_1,w_2)=(1,2,-3)$ with all other entries zero has $T\ne0$. The defect
therefore vanishes for every state if and only if $3\lambda-2=0$, so
$\lambda=\tfrac23$ is the unique energy-preserving member of the family, and
\eqref{eq:family} becomes the stencil \eqref{eq:skewsym} used in
\cref{sec:prob1,sec:prob2,sec:prob3}. \qed

The stencil is classical. Equivalently
\[
  \bigl[\Hm_F(\qq\otimes\qq)\bigr]_i
  =-\frac{(w_{i+1}+w_i+w_{i-1})(w_{i+1}-w_{i-1})}{6\Delta x} ,
\]
and the energy argument that selects it is that of Aref and
Daripa~\cite{ArefDaripa1984}. The same three-point average is the nonlinear
term of the Korteweg--de Vries scheme of Zabusky and
Kruskal~\cite{ZabuskyKruskal1965}, which \cref{sec:prob2} uses.

\section{A worked example}
\label{app:toy}

We illustrate the objects of \cref{sec:ep} and \cref{sec:bayes} on the
smallest non-trivial case $r=2$, $p=1$. By \cref{cor:compact} the quadratic part
of an EP model has $(r^3-r)/3=2$ free parameters; the general EP quadratic term
in two dimensions is
\begin{equation}
  \Hhat(\q\otimes\q)=(a\hat q_1+b\hat q_2)\begin{bmatrix}\hat q_2\\-\hat q_1\end{bmatrix},
  \qquad a,b\in\R,
  \label{eq:toyH}
\end{equation}
a rotation whose rate depends linearly on the state. Take $a=1$, $b=\tfrac12$ and
\begin{equation}
  \chat=\begin{bmatrix}0.2\\-0.1\end{bmatrix},\quad
  \Ahat=\begin{bmatrix}-1&2\\0&-3\end{bmatrix},\quad
  \Bhat=\begin{bmatrix}1\\0\end{bmatrix},\quad
  \Hhat=\begin{bmatrix}0&1&\tfrac12\\[2pt]-1&-\tfrac12&0\end{bmatrix},
  \label{eq:toyops}
\end{equation}
where the columns of $\Hhat$ hold the compact coefficients
$(h_{i,11},h_{i,12},h_{i,22})$ of \eqref{eq:compactH}.

\paragraph{Operator matrix and parameter vector.}
With $\bd(\q,u)=[\,1;\ \hat q_1;\ \hat q_2;\ \hat q_1^2;\ \hat q_1\hat q_2;\ \hat q_2^2;\ u\,]$ we have
$d=1+r+r(r+1)/2+p=7$ and
\begin{equation}
  \Ohat=[\,\chat\ \ \Ahat\ \ \Hhat\ \ \Bhat\,]
  =\begin{bmatrix}
    0.2&-1&2&0&1&0.5&1\\
    -0.1&0&-3&-1&-0.5&0&0
  \end{bmatrix}\in\R^{2\times7},
\end{equation}
and, stacking its rows, $\bth\in\R^{n}$ with $n=rd=14$,
\begin{equation}
  \bth=\bigl(\underbrace{0.2,\,-1,\,2,\,0,\,1,\,0.5,\,1}_{\ohat_1},\
            \underbrace{-0.1,\,0,\,-3,\,-1,\,-0.5,\,0,\,0}_{\ohat_2}\bigr).
\end{equation}
The quadratic entries sit in positions $4,5,6$ (row 1) and $11,12,13$ (row 2).

\paragraph{Constraint matrix.}
There are $n_c=\binom{r+2}{3}=4$ constraints, one per cubic monomial:
\begin{center}
\small
\begin{tabular}{@{}llc@{}}
\toprule
monomial & constraint \eqref{eq:compactC} & columns of $\bth$\\
\midrule
$\hat q_1^3$ & $h_{1,11}=0$ & $4$\\
$\hat q_1^2\hat q_2$ & $h_{1,12}+h_{2,11}=0$ & $5,11$\\
$\hat q_1\hat q_2^2$ & $h_{1,22}+h_{2,12}=0$ & $6,12$\\
$\hat q_2^3$ & $h_{2,22}=0$ & $13$\\
\bottomrule
\end{tabular}
\end{center}
so that, writing only the nonzero entries,
\begin{equation}
  \C=\begin{bmatrix}
   0&0&0&1&0&0&0&\ 0&0&0&0&0&0&0\\
   0&0&0&0&1&0&0&\ 0&0&0&1&0&0&0\\
   0&0&0&0&0&1&0&\ 0&0&0&0&1&0&0\\
   0&0&0&0&0&0&0&\ 0&0&0&0&0&1&0
  \end{bmatrix}\in\R^{4\times14}.
  \label{eq:toyC}
\end{equation}
The eight zero columns correspond to the entries of $\chat,\Ahat,\Bhat$, which
the constraint does not touch. Each column has at most one nonzero, so
$\C\C\T=\diag(1,2,2,1)$ and $\operatorname{rank}\C=4=n_c$, as in
\cref{cor:compact} and \cref{rem:attr-count}; the value $\nu_\ell=3$ does not
occur because the monomial $\hat q_a\hat q_b\hat q_c$ with distinct indices requires
$r\ge3$. One checks $\C\bth=\mathbf{0}$, and
$\dim\ker\C=14-4=10$: the eight unconstrained entries of $\chat,\Ahat,\Bhat$
together with the two parameters $a,b$ of \eqref{eq:toyH}.

\paragraph{Certificate.}
Here
$\sym(\Ahat)=\bigl[\begin{smallmatrix}-1&1\\1&-3\end{smallmatrix}\bigr]$ has
eigenvalues $-2\pm\sqrt2$, so $\alpha=2-\sqrt2\approx0.586>0$ and
\cref{thm:ISS} applies. With $\bar u=1$,
\begin{equation}
  K=\|\chat\|+\|\Bhat\|\,\bar u\approx1.224,
  \qquad R=K/\alpha\approx2.089 ,
\end{equation}
so every trajectory, from every initial condition and for every input with
$\|u\|_{L^\infty}\le1$, enters and remains in the ball of radius $2.089$. For
illustration, integrating with $u(t)=\sin(3t)$ from $\q_0=\begin{bmatrix} 0 & 0 \end{bmatrix}\T$, $\begin{bmatrix}5 & -4\end{bmatrix}\T$ and
$\begin{bmatrix}50 & 50\end{bmatrix}\T$ gives $\|\q(60)\|\approx0.19$ in all three cases, with suprema
$0.50$, $6.40$ and $70.71$ along the way: the bound is global, and, as expected
from the Cauchy--Schwarz and Young steps, conservative.

\end{document}